\documentclass[fleqn,usenatbib]{mnras}

\usepackage{newtxtext,newtxmath}

\usepackage[T1]{fontenc}

\DeclareRobustCommand{\VAN}[3]{#2}
\let\VANthebibliography\thebibliography
\def\thebibliography{\DeclareRobustCommand{\VAN}[3]{##3}\VANthebibliography}

\usepackage{graphicx}
\usepackage{amsmath}
\usepackage{colortbl}
\usepackage{placeins}

\title[LMC perturbations in the MW halo]{LMC-induced Perturbations in the Milky Way Halo II: Bridging Field-level Inference and Summary-level Simulation-Based Inference}

\author[Sheng et al.]{
Yanjun Sheng,$^{1}$\thanks{E-mail: Yanjun.Sheng@anu.edu.au}
Yuan-Sen Ting,$^{2,3,4}$
Tomasz Różański,$^{1}$
Xiang-Xiang Xue,$^{5,6}$
Roland M. Crocker,$^{1}$
\newauthor
Jiashu Pan$^{7}$
\\
$^{1}$Research School of Astronomy and Astrophysics, Australian National University, Canberra, ACT 2611, Australia\\
$^{2}$Department of Astronomy, The Ohio State University, Columbus, OH 43210, USA\\
$^{3}$Center for Cosmology and AstroParticle Physics (CCAPP), The Ohio State University, Columbus, OH 43210, USA\\
$^{4}$Max-Planck-Institut für Astronomie, Königstuhl 17, D-69117 Heidelberg, Germany\\
$^{5}$National Astronomical Observatories, Chinese Academy of Sciences, Beijing 100101, People’s Republic of China\\
$^{6}$Institute for Frontiers in Astronomy and Astrophysics, Beijing Normal University, Beijing 102206, People’s Republic of China\\
$^{7}$School of Engineering, Westlake University, Hangzhou, Zhejiang, 310030, People’s Republic of China
}

\date{Accepted XXX. Received YYY; in original form ZZZ}

\pubyear{\the\year{}}

\begin{document}
\label{firstpage}
\pagerange{\pageref{firstpage}--\pageref{lastpage}}
\maketitle

\begin{abstract}

The gravitational interaction between the Milky Way (MW) and the Large Magellanic Cloud (LMC) drives the outer halo into dynamical disequilibrium, imprinting the masses and structural parameters of both galaxies onto the 6D phase-space distribution of halo tracers. This signal has been characterised with summary statistics ranging from low-order velocity moments to basis function expansions, yet how much information these summaries discard, and whether they are complementary, remains unclear. We address these questions by comparing a likelihood of the halo phase-space distribution (field-level) with physically interpretable summaries for constraining $(M_{\mathrm{MW}}, M_{\mathrm{LMC}}, c, q)$, where $c$ and $q$ are the MW halo concentration and flattening. A Conditional Flow Matching (CFM) model trained on the HaloDance $N$-body suite provides an exact likelihood at a held-out fiducial point; for 5,000 tracers in $30$--$120$~kpc it tightens marginal constraints by factors of $2.5$--$9.9$ over an all-sky velocity-moment forecast. To narrow this gap, we construct four parameter-sensitive summaries via BFE+MOPED that capture the angular structure of the halo density and velocity and contain information complementary to the velocity moments. Combining the two sets of summaries tightens the marginal constraints by up to $15$ per cent relative to the angular summaries alone, and by $30$--$71$ per cent relative to the velocity moments alone, although the resulting constraints remain $1.3$--$2.9$ times broader than those from the full phase-space benchmark. We thus establish a physically interpretable summary-level route for applying this inference pipeline to future observations, while the field-level benchmark measures the information available for further improvement.

\end{abstract}


\section{Introduction}

The Large Magellanic Cloud (LMC) is the most massive satellite galaxy of the Milky Way (MW), with a virial mass estimated at $0.8$--$2.5 \times 10^{11}~\mathrm{M}_{\odot}$, representing 10--25 per cent of the MW's total mass \citep{BlandHawthorn2016,Shipp2021,Fushimi2024}. Independent constraints on the LMC mass come from its rotation curve \citep{vanderMarel2014}, three-dimensional kinematics of its globular clusters \citep{Watkins2024}, deflections of MW stellar streams \citep{Erkal2019,Shipp2021,Vasiliev2021,Koposov2023}, the cosmological timing argument \citep{Penarrubia2016,Chamberlain2023}, and MW halo stars entrained in the LMC's gravitational wake \citep{Fushimi2024,Cavieres2025}. The MW--LMC interaction produces dynamical signatures whose amplitudes depend on the mass distributions of both systems \citep{Gomez2015,Laporte2018b,Laporte2018a,Garavito-Camargo2019,Petersen2020,Erkal2021,Vasiliev2023,Kravtsov2024,Sheng2024,Sheng2025,Brooks2025,Brooks2026b,Brooks2026a,DarraghFord2025}.

The orbital history of the LMC has been fundamentally revised over the past two decades. Hubble Space Telescope proper motions, which show the LMC's tangential velocity approaching the MW escape speed \citep{Kallivayalil2006,Kallivayalil2013,Besla2007,Besla2018}, replaced the once-canonical picture of short-period, multi-orbit trajectories \citep[e.g.,][]{Tremaine1976,Murai1980,Lin1982} with a first-infall scenario in which the LMC is near pericentre at $\sim$50~kpc moving faster than 300~km~s$^{-1}$ \citep{Kallivayalil2013,Vasiliev2021,Sheng2024,Lucchini2025}; a second passage $\gtrsim$100~kpc some 5--10~Gyr ago nonetheless remains viable given uncertainties in the MW potential \citep{Vasiliev2024,Suzuki2026}. Independent of orbital history, the LMC's infall produces two principal dynamical effects in the MW halo: (i) a reflex motion of the inner Galaxy relative to the outer halo, manifesting as a dipole in the stellar density and radial velocity distribution \citep{Petersen2020,Petersen2021,Erkal2021}, and (ii) a dynamical friction wake, which is a localised overdensity of dark matter and stars trailing the LMC's past trajectory \citep{Chandrasekhar1943,Garavito-Camargo2019,Tamfal2021,Foote2023}. These effects depend not only on the mass of the LMC but also on the shape, concentration, and velocity anisotropy of the MW halo, making the inference problem intrinsically coupled.

Observations have established both signatures in the MW outer halo. The reflex motion appears as a systematic Galactocentric velocity dipole in halo tracers beyond 50~kpc \citep{Erkal2021, Petersen2021, Conroy2021, Bystrom2025, Chandra2025, Li2026b}, while the dynamical friction wake appears as a localized stellar overdensity trailing the LMC's past orbit, including the Pisces Plume \citep{Belokurov2019, Conroy2021, Fushimi2024, Cavieres2025}. These detections show that the MW's outer halo is far from equilibrium and that its 6D phase-space structure carries a coupled imprint of the LMC's gravitational influence.

On the theoretical side, idealised $N$-body simulations have characterised both the local wake and the global density dipole, and used these signatures to bracket the LMC mass \citep{Gomez2015,Garavito-Camargo2019,GaravitoCamargo2021,Tamfal2021,Foote2023, Vasiliev2023, Sheng2024}. Cosmological zoom-in runs confirm that the reflex motion and wake signatures persist in a full $\Lambda$CDM context with realistic merger histories \citep{DarraghFord2025}, and dedicated $N$-body suites have begun to systematically explore the multi-dimensional parameter space of MW mass, LMC mass, halo shape, and orbital configuration \citep{Sheng2025,Brooks2025,Garver2026}.

To extract physical insight from these simulations, multipole expansions of the density and potential fields (typically implemented as biorthogonal basis function expansions, or BFEs) decompose the perturbed halo into distinct angular and radial channels, separating the $\ell = 1$ reflex motion from the $\ell \geq 2$ local wake \citep{GaravitoCamargo2021,Lilleengen2023,DarraghFord2025,Foote2026}. The same biorthogonal projection underpins the matrix method of linear response theory, which has shown analytically that the reflex motion depends solely on the MW potential while the wake is sensitive to the halo velocity anisotropy \citep{Rozier2022}. Spherical harmonic expansions of the stellar velocity field offer a complementary route, applied to simulations \citep{Cunningham2020}. These frameworks together show that the MW's response to the LMC decomposes into interpretable multipole components, each encoding different aspects of the underlying mass distributions and orbital configuration of halo stars.

Two limitations remain. First, many analyses compare data with a small number of fiducial simulations rather than sampling the multi-dimensional parameter posterior. Second, the summaries used to compress this signal, most commonly low-order velocity moments (all-sky mean velocities and dispersions in broad radial shells), but also basis function expansions of the density and velocity fields, are adopted without knowing how much phase-space information each retains or whether different summaries are complementary. Low-order velocity moments in particular are interpretable but discard the localized structure of the reflex motion and wake. This gap has persisted partly because no absolute upper limit on the phase-space information was available against which any summary could be measured. The HaloDance suite \citep{Sheng2025} and neural posterior estimation \citep{Brooks2026b} address the parameter-space search, but still rely on low-order velocity moments without a benchmark for the total information content.

In this work we use field-level inference as a benchmark for the information content of the 6D phase-space distribution of halo tracers, and compare it against summary statistics whose physical channels can be inspected. We train a Conditional Flow Matching (CFM) generative model directly on the raw particle coordinates of the HaloDance simulations, this yields a likelihood for a tracer catalogue at any proposed parameter point and defines a reference information limit for this simulation setup. We then quantify how much of that information is retained by standard velocity moments and by a multipole BFE of the halo density and velocity fields compressed with the Multiple Optimised Parameter Estimation and Data compression (MOPED) algorithm, and use a variational mutual-information analysis to show that the two summary sets are complementary. Combining them yields our primary parameter constraints, while comparison against the field-level likelihood bounds how much information the interpretable summaries leave unrecovered.

\section{Datasets and emulation}
\label{sec:data_emulation}

The inference setup combines the HaloDance suite of high-resolution $N$-body simulations, which define the MW--LMC parameter space, with a Conditional Flow Matching (CFM) generative model trained on them. The CFM interpolates between discrete simulation points and provides a likelihood for 6D phase-space samples.

\subsection{HaloDance simulation suite}
\label{sec:halodance}

The analysis is built on the HaloDance suite of $N$-body simulations, which models the gravitational interaction between the MW and LMC and the resulting perturbation to the MW stellar halo. The MW model consists of an NFW dark matter halo \citep{NFW}, a Miyamoto--Nagai stellar disk \citep{Miyamoto}, and a Hernquist bulge \citep{Hernquist1990}. The LMC is modelled as a spherical Hernquist dark matter halo. Initial galaxy models are generated with the \textsc{galic} code \citep{Yurin2014} and evolved in isolation for 3 Gyr with \textsc{gadget-4} \citep{Springel2021} to ensure dynamical stability before the interaction begins.

The suite comprises 2,848 high-resolution simulations, each with $10^{7}$ particles and a particle mass of $10^{5}~\mathrm{M}_{\odot}$. The simulations span a four-dimensional parameter space using Latin Hypercube Sampling:
\begin{itemize}
    \item MW virial mass, $M_{\mathrm{MW}} \in [0.5, 2.0] \times 10^{12}~\mathrm{M}_{\odot}$ (fiducial: $0.7 \times 10^{12}~\mathrm{M}_{\odot}$),
    \item LMC virial mass, $M_{\mathrm{LMC}} \in [0.7, 2.1] \times 10^{11}~\mathrm{M}_{\odot}$ (fiducial: $1.5 \times 10^{11}~\mathrm{M}_{\odot}$),
    \item MW halo concentration, $c \in [5, 15]$ (fiducial: $9.415$),
    \item MW halo flattening, $q \in [0.5, 1.5]$ (fiducial: $1.0$).
\end{itemize}
We consider two tracer velocity anisotropy profiles $\beta$:
\begin{itemize}
    \item Isotropic, $\beta = 0$ (1,848 models; adopted as fiducial),
    \item Radially varying, $\beta(r) = -0.15 - 0.2\,\alpha(r)$ (1,000 models), where $\alpha(r) = d\ln\rho/d\ln r$ is the local density slope \citep{Hansen2006}.
\end{itemize}
The primary inference is conditional on the isotropic prescription, while Section~\ref{sec:summary_validation} quantifies the parameter biases that result when the isotropic-trained summary likelihood is applied to the radially varying suite. Only parameter combinations in which the LMC completes a single pericentric passage and reaches an apocentre beyond the MW's virial radius within the last 5 Gyr are retained, ensuring all models follow a first-infall scenario without imposing it by hand. Initial orbital coordinates for each simulation are determined via a multi-layer perceptron neural network trained to map initial phase-space coordinates to the observed present-day configuration \citep{Kallivayalil2013}. Each system is evolved for 2 Gyr with \textsc{gadget-4}.

For each simulation, we extract the 6D phase-space coordinates of dark matter particles within 30--120~kpc of the MW centre (i.e., the disk centre) as the raw data for our inference pipeline. We refer the reader to \cite{Sheng2025} for a detailed description of the simulation setup, parameter grid, and convergence tests.

\subsection{Conditional Flow Matching model}
\label{sec:cfm}

Conditional Flow Matching bridges normalising flows \citep{JimenezRezende2015,Durkan2019} and diffusion models \citep{Sohl-Dickstein2015,Ho2020,Song2020}. CFM regresses a neural network to the velocity field that transports probability mass from a simple base distribution to the target data distribution along continuous paths, while retaining the exact likelihood computation of continuous normalising flows. In astrophysics, flow matching has recently been applied to galaxy property inference from imaging \citep{Yunus2025} and inference of the Galactic potential from stellar streams \citep{Viterbo2026}. Here the CFM model learns the conditional phase-space distribution of MW halo particles, $p(\mathbf{x} \mid \boldsymbol{\theta})$, where $\mathbf{x}$ denotes the 6D phase-space vector. Specifically, CFM parameterises a time-dependent vector field with a neural network to define a continuous trajectory mapping a simple 6D Gaussian base distribution to the parameter-dependent phase-space distribution of the halo tracers. Because this mapping is continuous and invertible, it allows both mock stellar distributions to be generated by flowing noise particles forward and exact probability densities to be evaluated for observed stellar phase-space coordinates under the model.

\begin{figure*}
    \centering
    \includegraphics[width=0.86\textwidth]{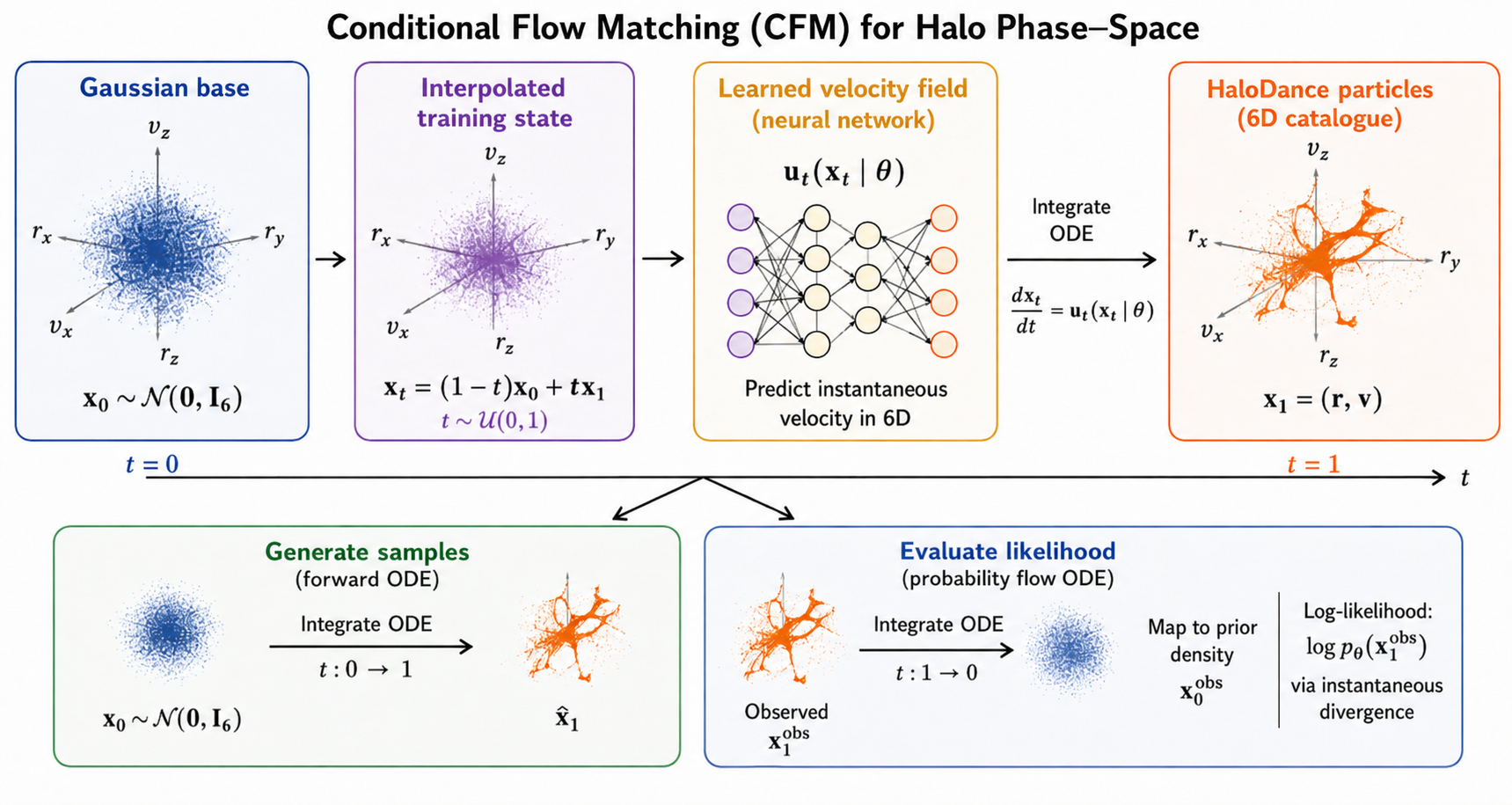}
    \caption{Schematic of the conditional flow matching setup. Gaussian base particles $\mathbf{x}_0$ are connected to HaloDance phase-space particles $\mathbf{x}_1=(\mathbf{r},\mathbf{v})$ through interpolated states $\mathbf{x}_t$. The network learns the conditional velocity field $\mathbf{u}_t(\mathbf{x}_t\mid\boldsymbol{\theta})$, forward integration generates mock halo tracers, while backward integration maps observed 6D catalogues to the base distribution for likelihood evaluation.}
    \label{fig:cfm_schedule}
\end{figure*}

Figure~\ref{fig:cfm_schedule} summarises this construction, from the Gaussian base particles to HaloDance phase-space particles and the forward and backward probability-flow operations. The learned velocity field $\mathbf{u}_t(\mathbf{x} \mid \boldsymbol{\theta})$ is conditioned on the MW--LMC parameters $\boldsymbol{\theta} = (M_{\mathrm{MW}}, M_{\mathrm{LMC}}, c, q)$, and its input is a 6D phase-space vector $\mathbf{x} = (x, y, z, v_x, v_y, v_z)$ representing position and velocity in Galactocentric coordinates, augmented with radial features to provide rotation-invariant spatial information.

The model is trained on a 90 per cent split of the isotropic HaloDance simulations, with 10 per cent held out for validation. The training process uses optimal-transport conditional flow matching (OT-CFM) to learn a straight-line probability path mapping Gaussian noise to the physical phase space. To ensure training stability and accuracy, we implement a curriculum weighting over the parameter space. Once trained, the likelihood of a given particle sample is computed using the continuous change-of-variables formula. The full technical specifications of the neural network architecture, parameter embeddings, training hyperparameters, and probability flow integration are detailed in Appendix~\ref{app:cfm_details}.

\begin{figure*}
    \centering
	\includegraphics[width=0.7\textwidth]{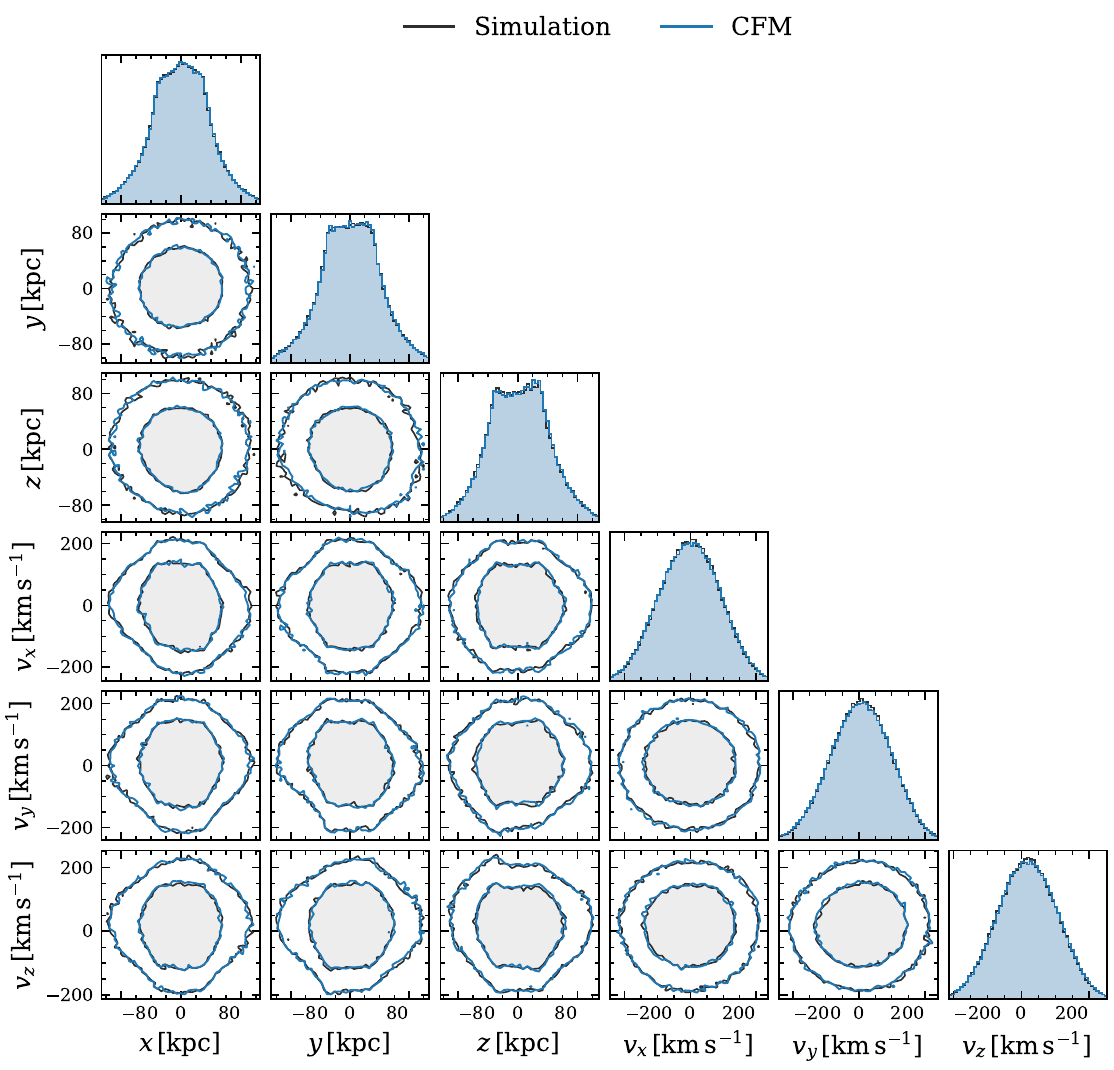}
    \caption{Pairwise comparison of the six Galactocentric phase-space coordinates for CFM-generated particles (blue) and particles from a held-out fiducial simulation (grey), both restricted to $30$--$120$~kpc. The fiducial point is $(M_{\mathrm{MW}},M_{\mathrm{LMC}},c,q)=(0.7\times10^{12}\,\mathrm{M}_\odot,1.5\times10^{11}\,\mathrm{M}_\odot,9.415,1.0)$ with $\beta(r)=0$. The diagonal panels show the one-dimensional marginals of $x$, $y$, $z$ (kpc) and $v_x$, $v_y$, $v_z$ (km~s$^{-1}$), and the lower triangle shows the pairwise distributions. The two samples overlap across all panels, showing that the CFM matches both the one-dimensional marginals and the pairwise structure at a parameter point outside its training set.}
    \label{fig:cfm_generation}
\end{figure*}

We first verify that the CFM model generalises beyond its training set. Figure~\ref{fig:cfm_generation} compares the phase-space distribution of the six coordinates $(x, y, z, v_x, v_y, v_z)$ for $5 \times 10^5$ CFM-generated particles (blue) and an equal number of particles from a held-out fiducial simulation ($M_{\mathrm{MW}}=0.7\times 10^{12}\,\mathrm{M}_\odot$, $M_{\mathrm{LMC}}=1.5\times 10^{11}\,\mathrm{M}_\odot$, $c=9.415$, $q=1.0$, $\beta(r)=0$) that was not included in HaloDance training (grey). Both samples are restricted to the radial range 30--120~kpc, and CFM sampling uses a fixed-step RK4 ODE solver with 128 steps. The one-dimensional marginals on the diagonal and the pairwise distributions below them overlap between the two samples, with no visible systematic shifts or shape discrepancies, showing that the CFM matches the bulk phase-space distribution at a parameter point outside its training set. This check probes only low-order structure, and the benchmark in Section~\ref{sec:cfm_full_phase_space_benchmark} shows that the CFM also captures the localised, higher-order structure, such as the LMC wake, that carries much of the parameter information.

The same model can also assign a likelihood to an observed particle catalogue at any proposed $\boldsymbol{\theta}$. For a single particle, the CFM log-likelihood follows from the continuous normalising-flow interpretation and is computed by integrating the probability flow ODE backward from the data to the Gaussian base distribution, keeping track of the instantaneous change of variables. For a sample of $N$ independent particles, the total log-likelihood is the sum of the per-particle log-likelihoods. The full mathematical formulation of the change-of-variables ODE, coordinate normalisation, divergence estimation via the Hutchinson trace estimator, and the ODE solver settings are detailed in Appendix~\ref{app:cfm_details}. This likelihood is what allows the CFM to serve as a benchmark for the information content of the raw 6D phase-space distribution.

\subsection{Field-level inference benchmark}
\label{sec:cfm_full_phase_space_benchmark}

We evaluate the CFM log-likelihood on a grid around one held-out fiducial simulation using the following setup.

\begin{itemize}
    \item Fiducial point, $\boldsymbol{\theta}_\star = (M_{\mathrm{MW}}, M_{\mathrm{LMC}}, c, q) = (0.70\times 10^{12}\,\mathrm{M}_\odot,\;1.50\times 10^{11}\,\mathrm{M}_\odot,\;9.415,\;1.0)$.
    \item Grid bounds, $M_{\mathrm{MW}}\in[0.68,\,0.73]\times 10^{12}\,\mathrm{M}_\odot$, $M_{\mathrm{LMC}}\in[1.44,\,1.68]\times 10^{11}\,\mathrm{M}_\odot$, $c\in[8.0,\,10.6]$, and $q\in[0.97,\,1.03]$.
    \item Mock data, $N=5{,}000$ particles drawn from the held-out fiducial simulation and restricted to 30--120~kpc.
\end{itemize}

Because the benchmark is local and four-dimensional, we evaluate the likelihood on an 11-point grid per parameter ($11^4 = 14{,}641$ points), which yields the normalised posterior without sampler-convergence choices and parallelises trivially across GPUs. At each grid point we sum the per-particle log-likelihoods, $\log\mathcal{L}(\boldsymbol{\theta})=\sum_{i=1}^{N}\log p(\mathbf{x}_i\mid\boldsymbol{\theta})$. With a uniform prior over the grid, the discrete posterior is $p(\boldsymbol{\theta}\mid\mathbf{x})\propto\exp[\log\mathcal{L}(\boldsymbol{\theta})]$, normalised by the Riemann sum over the grid. 

We compare this particle-level posterior with a Fisher forecast from the velocity moments used in \cite{Sheng2025}. The moment vector $\mathbf{v}$ has 15 components: the mean velocities $\langle v_r\rangle$ and $\langle v_b\rangle$ together with the dispersions $\sigma_{v_r}$, $\sigma_{v_b}$, and $\sigma_{v_\ell}$, measured in each of three Galactocentric distance shells (30--60, 60--90, and 90--120~kpc). Here $v_r$, $v_b$, and $v_\ell$ denote the radial and the two tangential (Galactic latitude and longitude) velocity components. This baseline compresses each halo to all-sky first and second velocity moments in radial bins, without using the angular structure of the density field, the wake, or the full 6D particle distribution.

Assuming a Gaussian likelihood for the summary vector $\mathbf{v}$, the Fisher information matrix $\mathbf{F} \in \mathbb{R}^{P \times P}$ for the $P=4$ model parameters is defined at the fiducial parameter point $\boldsymbol{\theta}_\star$ as:
\begin{equation}
F_{ij} = \frac{\partial \boldsymbol{\mu}_{\mathbf{v}}^\top}{\partial \theta_i} \mathbf{C}_{\mathbf{v}}^{-1} \frac{\partial \boldsymbol{\mu}_{\mathbf{v}}}{\partial \theta_j},
\label{eq:fisher_moments}
\end{equation}
where $\boldsymbol{\mu}_{\mathbf{v}}(\boldsymbol{\theta}) \equiv \langle\mathbf{v}\rangle_{\boldsymbol{\theta}}$ is the mean summary vector at parameter point $\boldsymbol{\theta}$, and $\mathbf{C}_{\mathbf{v}}$ is the summary covariance matrix. The covariance matrix $\mathbf{C}_{\mathbf{v}}$ is estimated from $N_{\rm real} = 2{,}000$ independent realisations of the fiducial simulation:
\begin{equation}
\mathbf{C}_{\mathbf{v}} = \frac{1}{N_{\rm real}-1} \sum_{k=1}^{N_{\rm real}} (\mathbf{v}^{(k)} - \bar{\mathbf{v}})(\mathbf{v}^{(k)} - \bar{\mathbf{v}})^\top,
\label{eq:covariance_estimation}
\end{equation}
where $\mathbf{v}^{(k)}$ is the summary vector of the $k$-th realisation and $\bar{\mathbf{v}}$ is the sample mean. The parameter derivatives are computed by central finite differences:
\begin{equation}
\frac{\partial \boldsymbol{\mu}_{\mathbf{v}}}{\partial \theta_i} \approx \frac{\boldsymbol{\mu}_{\mathbf{v}}(\boldsymbol{\theta}_\star + \Delta\theta_i\hat{\mathbf{e}}_i) - \boldsymbol{\mu}_{\mathbf{v}}(\boldsymbol{\theta}_\star - \Delta\theta_i\hat{\mathbf{e}}_i)}{2\Delta\theta_i},
\label{eq:derivative_estimation}
\end{equation}
where $\hat{\mathbf{e}}_i$ is the unit vector along the $i$-th parameter direction, $\Delta\theta_i$ is a small step size, and the mean summary vectors $\boldsymbol{\mu}_{\mathbf{v}}(\boldsymbol{\theta}_\star \pm \Delta\theta_i\hat{\mathbf{e}}_i)$ are evaluated by averaging over $N_{\rm deriv} = 100$ independent paired realisations per parameter direction. Under this Gaussian summary-likelihood approximation, the inverse Fisher matrix gives the Cram\'er--Rao bound on the parameter covariance, and the forecast marginalised $1\sigma$ constraints follow as $\sigma_{\alpha} = \sqrt{[F^{-1}]_{\alpha\alpha}}$.

\begin{figure}
	\includegraphics[width=\columnwidth]{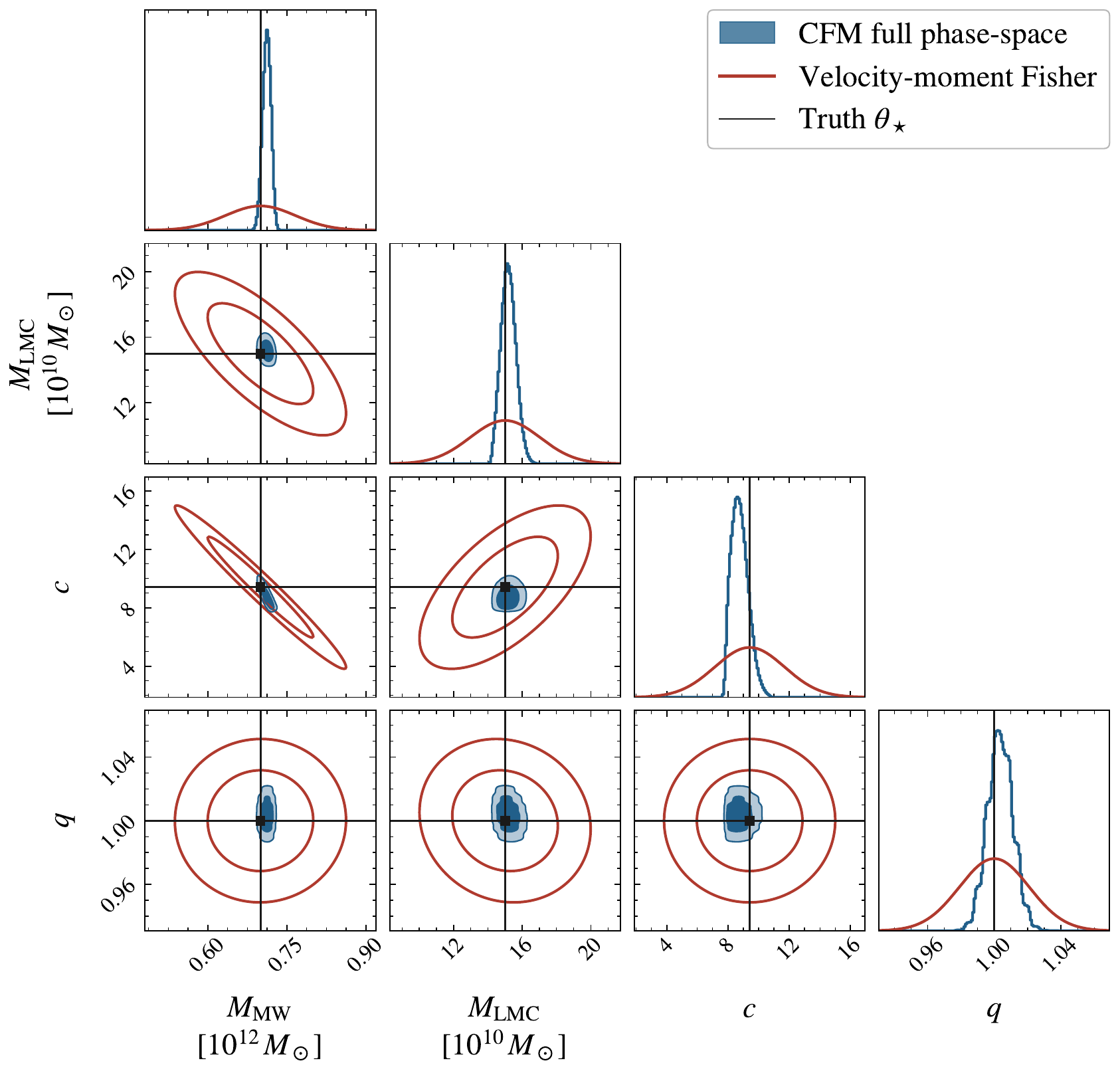}
    \caption{CFM-based full-phase-space likelihood benchmark for the held-out fiducial simulation at $\boldsymbol{\theta}_\star$ with $\beta(r)=0$. The corner plot overlays two contour sets in the same parameter planes: the CFM full-phase-space posterior from the likelihood grid for $N=5{,}000$ particles in $30$--$120$~kpc (blue, filled), and the velocity-moment Fisher forecast (red, Gaussian contours). Both sets show the 68 and 95 per cent credible regions, and black lines mark the true parameters $\boldsymbol{\theta}_\star$.}
    \label{fig:cfm_fiducial_benchmark}
\end{figure}

\begin{table}
\centering
\caption{Fiducial marginal $1\sigma$ constraints from the full phase-space CFM posterior and the velocity-moment Fisher forecast.}
\label{tab:cfm_velocity_constraint_comparison}
\begin{tabular}{lccc}
\hline
Parameter & CFM & Velocity moments & $\sigma_{\mathrm{vel}}/\sigma_{\mathrm{CFM}}$ \\
\hline
$M_{\mathrm{MW}}$ & 0.066 & 0.655 & 9.9 \\
$M_{\mathrm{LMC}}$ & 0.36 & 2.02 & 5.6 \\
$c$ & 0.52 & 2.26 & 4.3 \\
$q$ & 0.0084 & 0.021 & 2.5 \\
\hline
\end{tabular}
\par
\medskip
\begin{flushleft}
{\small \textbf{Notes.} Mass constraints are in units of $10^{11}\,\mathrm{M}_\odot$ for $M_{\mathrm{MW}}$ and $10^{10}\,\mathrm{M}_\odot$ for $M_{\mathrm{LMC}}$; $c$ and $q$ are dimensionless. The final column reports the ratio of the uncertainties $\sigma_{\mathrm{vel}}/\sigma_{\mathrm{CFM}}$. The CFM constraints are half the 16th to 84th percentile range of the one-dimensional marginal posteriors.}
\end{flushleft}
\end{table}

Figure~\ref{fig:cfm_fiducial_benchmark} shows the one- and two-dimensional marginal posteriors from the CFM grid search. The CFM contours are the credible regions of the grid posterior after smoothing the two-dimensional marginals with a Gaussian kernel of one-bin width, while the velocity-moment contours are analytic Gaussian ellipses at the matching radii. The fiducial truth $\boldsymbol{\theta}_\star$ lies within the 68 per cent contour in all panels, showing that the particle-level CFM likelihood recovers the input MW--LMC parameters in this test. Table~\ref{tab:cfm_velocity_constraint_comparison} gives the comparison to the velocity-moment Fisher forecast. The last column reports the ratio of the velocity-moment marginal uncertainty to the CFM marginal uncertainty, so it gives the constraint-tightening factor directly. The CFM posterior is tighter than the velocity-moment forecast by factors of $9.9$, $5.6$, $4.3$, and $2.5$ for $(M_{\mathrm{MW}}, M_{\mathrm{LMC}}, c, q)$, respectively. Thus, at this fiducial point, the raw catalogue carries information that is not captured by radial-bin means and dispersions alone.

Two limitations motivate the summary-level analysis that follows. First, evaluating the exact field-level likelihood requires a dense CFM grid and is computationally demanding, so this benchmark is restricted to the single fiducial point; validation across the broader parameter space instead uses the summary-level pipeline developed in Section~\ref{sec:compression}. The fiducial comparison is nevertheless not atypical: $\boldsymbol{\theta}_\star$ lies near the centre of the HaloDance ranges in $M_{\mathrm{LMC}}$, $c$, and $q$, and across all $185$ validation grids the summary-level posteriors preserve the same qualitative ordering, with both the BFE+MOPED and joint-summary posteriors tighter than the velocity-moment posteriors for every parameter, by factors of up to about two. The gain over velocity moments is therefore not tied to this particular fiducial choice. Second, the benchmark does not identify which structures carry that information. The neural transport field provides a likelihood for the catalogue, but it does not separate angular modes, radial scales, or velocity channels. Because tracer anisotropy, halo shape, selection effects, and wake modelling can affect these channels differently, the phase-space distribution must be compressed into summaries that can be mapped back to physical halo perturbations.

\section{BFE decomposition + MOPED compression}
\label{sec:compression}

The BFE+MOPED compression is applied to the phase-space distribution to retain information in summaries whose physical content can be inspected. This interpretability matters because simulation systematics can affect the reflex dipole, wake-like multipoles, first-moment velocities, and second-moment velocity channels differently. In the BFE representation, each coefficient, and any linear combination of coefficients, corresponds to a spatial and kinematic mode of the halo. The compression can therefore be traced back to the harmonic orders, radial scales, and velocity-moment channels that drive the constraints on each parameter.

\subsection{BFE decomposition of the phase space}
\label{sec:bfe_decomposition}

\begin{figure}
	\includegraphics[width=\columnwidth]{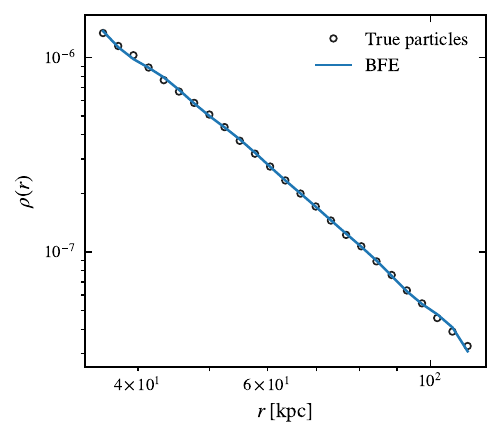}
    \caption{Spherically averaged density profile of particles drawn from the fiducial simulation within $30$--$120$~kpc. Open circles show the particle density estimate, and the solid blue line shows the BFE reconstruction with $n_{\max}=40$ and $\ell_{\max}=5$. The BFE traces the particle data across the full radial range, confirming that the adopted radial basis captures the scales relevant for inference.}
    \label{fig:bfe_density_profile}
\end{figure}

\begin{figure*}
	\includegraphics[width=\textwidth]{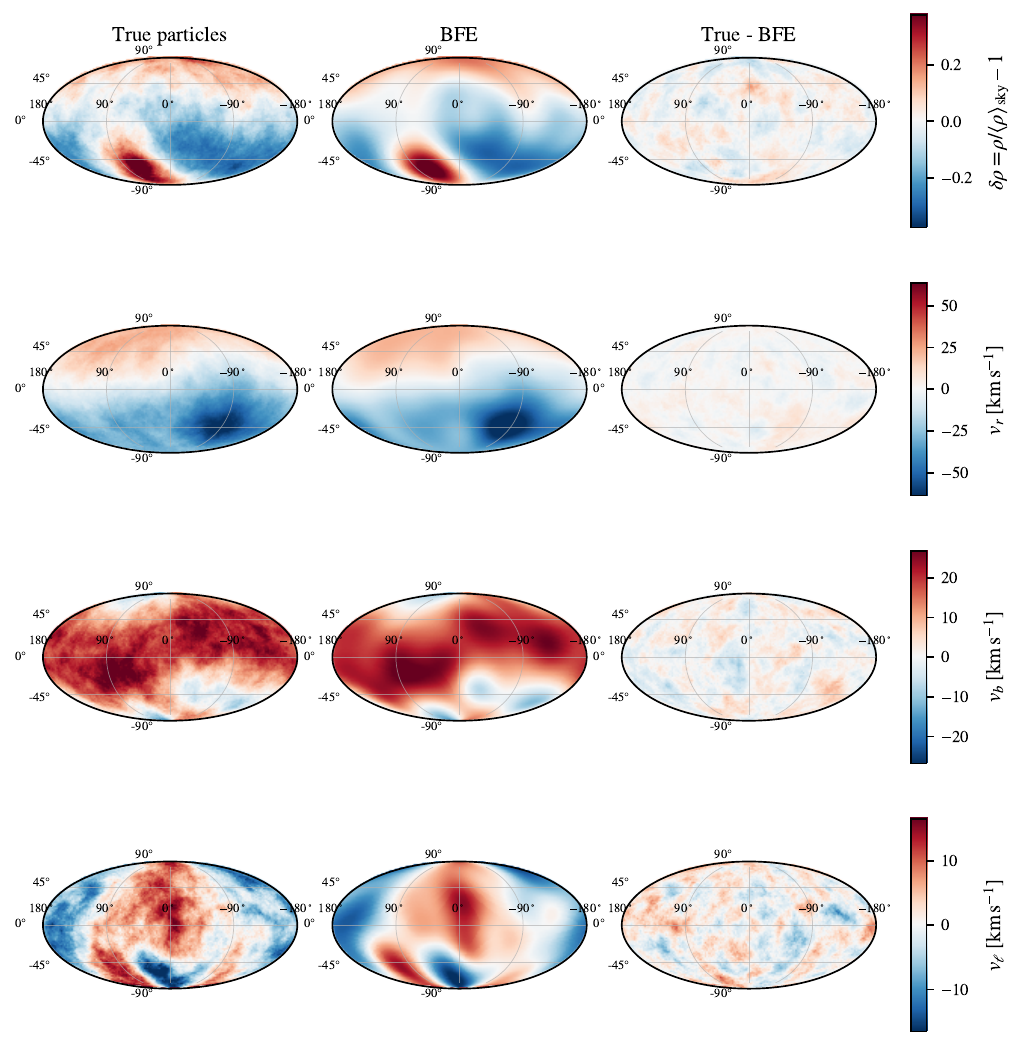}
    \caption{BFE reconstruction of the fiducial density and velocity fields in the $60$--$90$~kpc shell. Rows show the fractional density contrast $\delta\rho=\rho/\langle\rho\rangle_{\mathrm{sky}}-1$ and the Galactocentric velocities $v_r$, $v_b$, and $v_\ell$. Columns show the particle field, the BFE reconstruction with $n_{\max}=40$ and $\ell_{\max}=5$, and the residual. This figure checks that the BFE basis represents the large-scale density asymmetry, radial-velocity reflex dipole, and smaller angular velocity structure before MOPED compression is applied.}
    \label{fig:bfe_mollweide}
\end{figure*}

A BFE represents a self-gravitating system with paired density and potential basis functions. The term ``biorthogonal'' means that the density basis and the potential basis are orthogonal to each other as a pair. Projecting the particle distribution onto the potential basis gives the coefficient of the matching density basis function. This pairing matters because each density basis function and its potential partner satisfy Poisson's equation together.

We implement the expansion using the publicly available \textsc{EXP} code \citep{Petersen2025}. The density and potential fields of a discrete system of $N$ equal-mass particles are both expanded over the same amplitude coefficients $a_{\mu}$,
\begin{align}
\rho(\mathbf{x}) &= \sum_{\mu} a_{\mu}\,\rho_{\mu}(\mathbf{x}), \notag\\
\Phi(\mathbf{x}) &= \sum_{\mu} a_{\mu}\,\phi_{\mu}(\mathbf{x}), \notag\\
a_{\mu} &= \frac{1}{N} \sum_{i=1}^{N} \phi_{\mu}(\mathbf{x}_i),
\end{align}
where $\{\rho_{\mu}, \phi_{\mu}\}$ are biorthogonal density--potential pairs satisfying Poisson's equation $\nabla^{2}\phi_{\mu} = 4\pi G\,\rho_{\mu}$ and the biorthogonality condition $\int_{0}^{\infty} \phi_{\mu}\,\rho_{\nu}\,w(r)\,\mathrm{d}r = \delta_{\mu\nu}$, with $w(r) \equiv r^{2}/4\pi G$ \citep{DarraghFord2025}. In spherical coordinates, Poisson's equation is separable, so the basis functions factorize as
\begin{equation}
\rho_{\mu}(\mathbf{x}) = \rho_{n\ell}(r)\,Y_{\ell}^{m}(\theta,\phi), \qquad
\phi_{\mu}(\mathbf{x}) = \phi_{n\ell}(r)\,Y_{\ell}^{m}(\theta,\phi),
\end{equation}
with multi-index $\mu = (n, \ell, m)$. The spherical harmonics $Y_{\ell}^{m}$ carry the angular structure of each mode. The radial functions $\{\rho_{n\ell}(r), \phi_{n\ell}(r)\}$ are the eigenfunctions of the radial Sturm--Liouville problem set by Poisson's equation and the biorthogonality condition. In practice, \textsc{EXP} computes these radial functions for a user-specified input density profile.

The angular order $\ell$ gives each density coefficient a physical interpretation. The $\ell = 0$ terms encode the spherically symmetric mass distribution, or NFW monopole. The $\ell = 1$ terms encode the density dipole produced when the MW inner halo is displaced relative to the outer halo by the LMC's gravitational pull. Higher orders, $\ell \geq 2$, encode halo morphology beyond the dipole, including the LMC dynamical-friction wake. The $n = 0$ eigenfunction reproduces the input profile, and higher-$n$ eigenfunctions add radial nodes. In this study, we adopt an NFW profile with scale radius $r_{s} = 16$~kpc as the zeroth-order basis and expand to $\ell_{\max} = 5$ and $n_{\max} = 40$ radial orders over the radial range $0.5$--$300$~kpc.

To verify that these expansion orders can describe perturbed MW halos in simulation, we apply the BFE to a sample of $5 \times 10^{5}$ particles drawn from the fiducial simulation within $30$--$120$~kpc in Galactocentric coordinates, the same particle set used for the comparison in Figure~\ref{fig:cfm_generation}. Figure~\ref{fig:bfe_density_profile} shows that $n_{\max} = 40$ radial orders reconstruct the spherically averaged density profile of this sample. The BFE line traces through the particle density estimate across the full radial range, indicating that the basis captures the radial structure of the halo over the scales relevant for inference.

To go beyond the density distribution and capture the velocity structure, we extend the BFE to velocity-weighted projections. Since the BFE projection is linear in particle weights, replacing the mass weight $m_i$ with $m_i v_{k,i}$ yields coefficients that project the momentum density field $j_k(\mathbf{x}) \equiv \rho(\mathbf{x})\langle v_k \rangle(\mathbf{x})$ onto the same biorthogonal basis,
\begin{equation}
b_{\mu}^{k} = \frac{1}{N} \sum_{i=1}^{N} v_{k,i}\,\phi_{\mu}(\mathbf{x}_i),
\label{eq:bfe_momentum}
\end{equation}
for each Cartesian component $k \in \{x,y,z\}$. Here $\phi_{\mu}(\mathbf{x}_i)$ is evaluated at the particle position, while $v_{k,i}$ is only the velocity weight applied to that spatial projection. The expansion is therefore a velocity-weighted spatial BFE, not a basis expansion over velocity coordinates. The mean velocity field is then recovered point-wise as $\langle v_k \rangle(\mathbf{x}) = j_k(\mathbf{x})/\rho(\mathbf{x})$. Similarly, replacing weights with $m_i v_{k,i} v_{l,i}$ yields coefficients $c_{\mu}^{kl}$ for the second velocity-moment field $\rho\langle v_k v_l\rangle$, from which the velocity dispersion tensor $\sigma_{kl}^2(\mathbf{x}) = \langle v_k v_l\rangle(\mathbf{x}) - \langle v_k\rangle(\mathbf{x})\langle v_l\rangle(\mathbf{x})$ is obtained. For particles restricted to $30$--$120$~kpc, these projections give one density channel, three momentum-density channels, and six independent second-moment channels. Together they form the high-dimensional phase-space feature vector used for compression.

Figure~\ref{fig:bfe_mollweide} is a reconstruction test for the BFE representation, not an inference result. The left column shows the particle-estimated field, the middle column shows the BFE reconstruction, and the right column shows the particle field minus the BFE reconstruction. The density row contains the large-scale north--south asymmetry and the local wake. The radial-velocity row contains the strongest first-moment kinematic response, a coherent reflex dipole with amplitude $|v_r| \sim 25$~km~s$^{-1}$. The $v_b$ and $v_\ell$ rows have lower-amplitude angular structure. The residual maps are small compared with the large-scale patterns that are later compressed by MOPED. Across the three velocity components the absolute residual is comparable, about $2.9$~km~s$^{-1}$, and does not grow for the lower-amplitude $v_b$ and $v_\ell$ fields (with RMS amplitudes of $16.7$ and $7.3$~km~s$^{-1}$, against $26.2$~km~s$^{-1}$ for $v_r$). This component-independent residual reflects finite-sampling (shot) noise rather than an inadequate angular basis: the residual falls with particle number at close to the expected $1/\sqrt{N}$ rate (fitted slope magnitudes of $0.31$--$0.40$, with a small residual floor), whereas increasing $\ell_{\max}$ beyond $5$--$6$ does not improve the reconstruction.

\subsection{Linear compression with MOPED}
\label{sec:MOPED}

With the full $\ell_{\max}=5$, $n_{\max}=40$ BFE basis, the 10 channels contain $14{,}400$ active coefficients per halo ($36$ real harmonics per radial order and channel, after discarding the redundant $m=0$ sine terms of the raw storage); the adopted truncation at $n_{\max}=30$, chosen below by the Fisher-volume sweep, retains $10{,}800$. Even after this masking, the feature vector $\mathbf{f}$ is too high-dimensional for robust likelihood estimation from the available simulations, so we apply the MOPED algorithm, which reduces $\mathbf{f}$ to exactly $P=4$ linear combinations, one per parameter, while preserving the Fisher information of the original data. When the noise covariance matrix $\mathbf{C}$ of $\mathbf{f}$ is independent of the parameters, the Fisher matrix computed from the $P$ compressed summaries is identical to that from the full data vector.

In the original MOPED formulation the optimal weight vector for parameter $\theta_\alpha$ is $\mathbf{b}_\alpha \propto \mathbf{C}^{-1}\boldsymbol{\mu}_{,\alpha}$, where $\boldsymbol{\mu}_{,\alpha} \equiv \partial\langle\mathbf{f}\rangle/\partial\theta_\alpha$ is the gradient of the mean feature vector with respect to $\theta_\alpha$, and successive vectors $\{\mathbf{b}_\alpha\}$ are made mutually orthogonal by Gram--Schmidt construction (\cite{Heavens2000}, eq. 14).

We implement the MOPED projection using a standard three-step construction at the fiducial point $\boldsymbol{\theta}_\star$. First, BFE features are standardised using $2{,}000$ independent realisations at $\boldsymbol{\theta}_\star$, adopting a diagonal covariance approximation for the high-dimensional feature vector. Second, the parameter derivatives (scores) of the feature means are evaluated by central finite differences using realisations drawn from the trained CFM emulator. Finally, the projection matrix is constructed via QR-decomposition of the score matrix, which acts as a Gram--Schmidt orthogonalisation of the parameter-sensitive directions. The full mathematical derivation, standardisation procedure, and step-by-step algorithm are detailed in Appendix~\ref{app:moped_details}. 

We denote the resulting four-dimensional BFE+MOPED summary vector by $\mathbf{m}$. Its components are written as $s_{M_{\mathrm{MW}}}$, $s_{M_{\mathrm{LMC}}}$, $s_c$, and $s_q$, where each component is the MOPED projection optimised for the corresponding parameter direction at $\boldsymbol{\theta}_\star$. This notation separates the BFE+MOPED summaries $\mathbf{m}$ from the generic summary variable $\mathbf{s}$ used later in the mutual-information definition.

\begin{figure}
    \centering
    \includegraphics[width=\columnwidth]{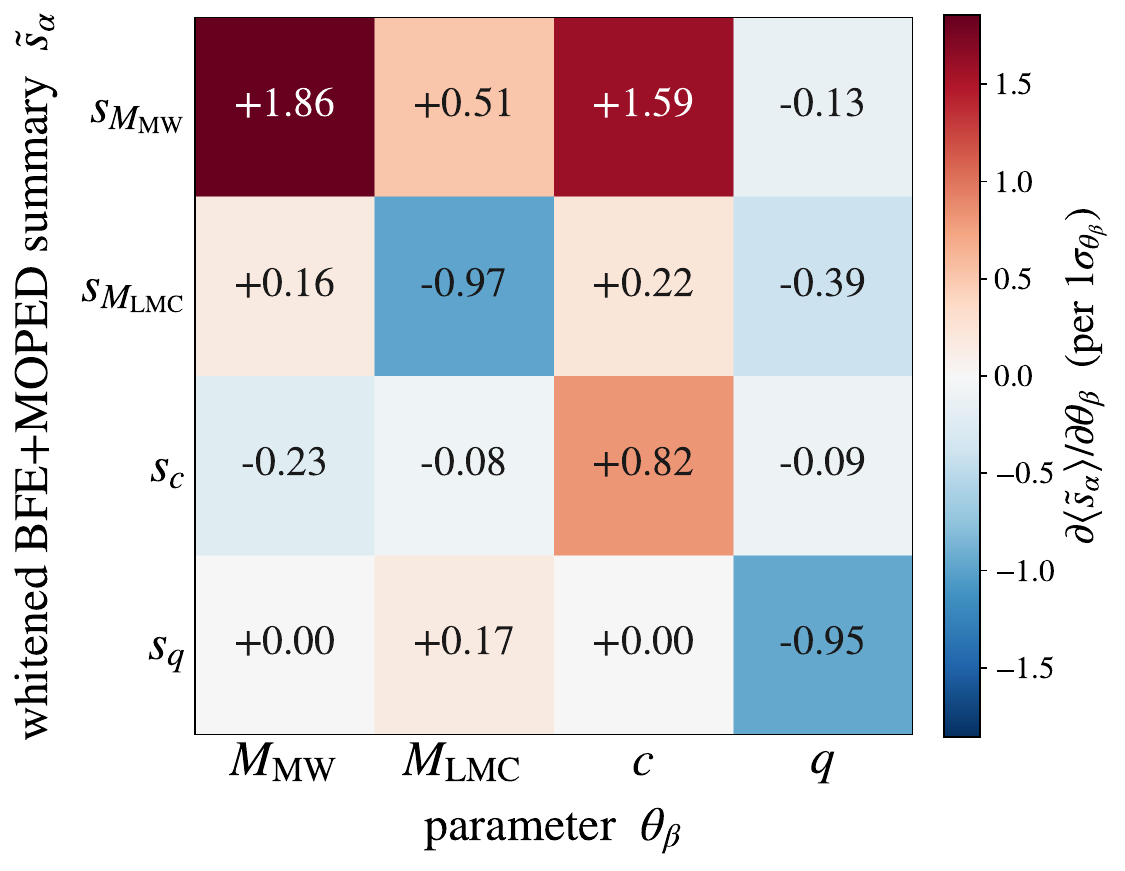}
    \caption{Response matrix of the four whitened BFE+MOPED summaries to the model parameters, $R_{\alpha\beta}=\partial\langle\tilde{s}_\alpha\rangle/\partial\theta_\beta$, evaluated at the fiducial point and shown in units of the response per $1\sigma$ shift of each parameter. Rows are the summaries $s_{M_{\mathrm{MW}}}$, $s_{M_{\mathrm{LMC}}}$, $s_c$, and $s_q$, and columns are the model parameters $(M_{\mathrm{MW}}, M_{\mathrm{LMC}}, c, q)$. The overall sign of each summary is conventional, so only relative signs within a row are meaningful. The matrix is diagonally dominant, so each summary tracks its intended parameter, and the largest off diagonal entries reflect the mass--concentration and mass--mass degeneracies.}
    \label{fig:moped_response}
\end{figure}

As a diagnostic of the fiducial MOPED construction, Figure~\ref{fig:moped_response} shows the response matrix $R_{\alpha\beta}=\partial\langle\tilde{s}_\alpha\rangle/\partial\theta_\beta$ of the whitened summaries, in units of the response per $1\sigma$ parameter shift. The matrix is diagonally dominant: each summary mainly tracks its intended parameter, and the $M_{\mathrm{LMC}}$, $c$, and $q$ summaries are cleanly isolated, with off-diagonal terms below about half of their diagonals. The two largest off-diagonal terms both appear in the MW-mass summary: its response to $c$ reflects the familiar NFW mass--concentration degeneracy, and its response to $M_{\mathrm{LMC}}$, about a quarter of its self-response and of the same sign, means a heavier LMC shifts the MW-mass summary in the same direction as a heavier MW. These couplings match the correlated mass constraints in the Fisher forecast and reflect physical degeneracies rather than shortcomings of the compression.

We use this fiducial transfer matrix for all BFE+MOPED summaries, so the channel-level physical decomposition should be interpreted as a local diagnostic. We assess the transfer-matrix stability using $10$ independent, non-fiducial anchor simulations drawn broadly across the prior volume, with $M_{\mathrm{MW}} \in [0.50, 1.80] \times 10^{12}~\mathrm{M}_\odot$, $M_{\mathrm{LMC}} \in [7.9, 18.9] \times 10^{10}~\mathrm{M}_\odot$, $c \in [6.2, 14.8]$, and $q \in [0.51, 1.31]$. At each anchor, we recompute the MOPED transfer matrix and compare its parameter-sensitive directions with the fiducial ones through their principal angles. The best-matched direction stays close to the fiducial one across the prior (principal cosine $0.70$--$0.84$), while the remaining directions rotate more, reaching principal angles of $64$--$79^\circ$. Resolved by parameter, the $c$ and $M_{\mathrm{MW}}$ directions are the most stable, with mean absolute cosines to the fiducial direction of $0.74$ and $0.68$, followed by $q$ at $0.64$, while the $M_{\mathrm{LMC}}$ direction is the least stable at $0.43$. The channel decomposition is therefore most reliable for $c$ and $M_{\mathrm{MW}}$, and weakest for $M_{\mathrm{LMC}}$.

In the exact Fisher-information limit, adding features cannot reduce the information in the data vector. In our finite-simulation implementation, however, the covariance and score derivatives are estimated from a limited number of realisations, so high-order radial modes can amplify estimation noise. We therefore truncate the feature vector to the first $n_{\max}$ radial orders per harmonic per channel before compression, choosing $n_{\max}$ to minimise the estimated joint Fisher posterior volume $\sqrt{\det F^{-1}}$ of the four MOPED summaries at the fiducial point.

\begin{figure*}
    \centering
    \includegraphics[width=0.82\textwidth]{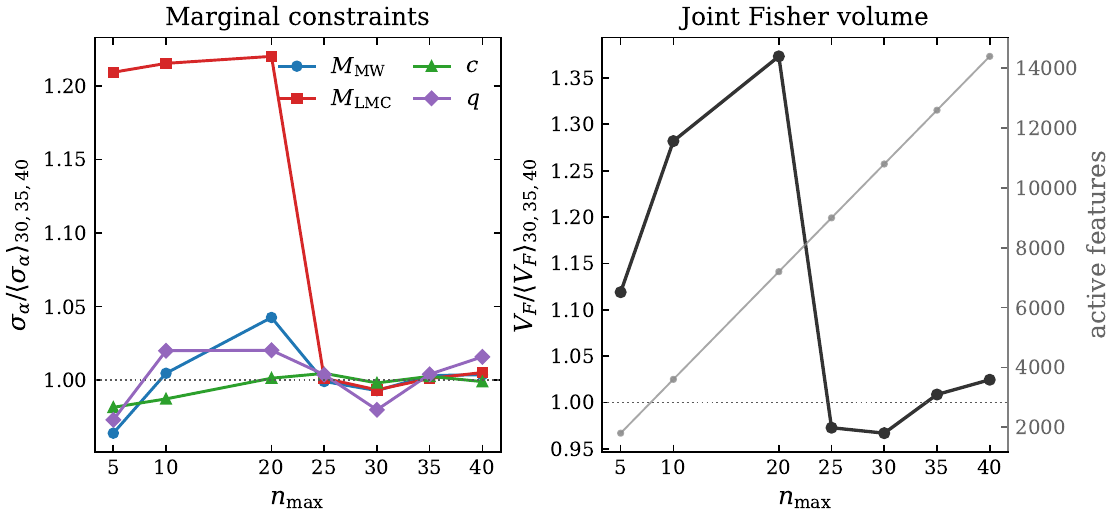}
    \caption{Radial truncation diagnostic for the BFE+MOPED compression. The left panel shows the four marginal constraints, each normalised by the mean over the plateau values $n_{\max}=30$, $35$, and $40$, so that $M_{\mathrm{MW}}$, $M_{\mathrm{LMC}}$, $c$, and $q$ appear together. The right panel shows the joint Fisher volume normalised the same way, with the number of active BFE features on the secondary axis. The joint volume reaches its minimum at $n_{\max}=30$, after which adding radial orders increases the active feature count without improving the forecast.}
    \label{fig:nmax_ankle}
\end{figure*}

We sweep $n_{\max}$ over the full range accessible from the $n_{\max}=40$ BFE basis. Figure~\ref{fig:nmax_ankle} summarises the sweep, normalising each marginal constraint and the joint Fisher volume by the mean over the plateau values $n_{\max}=30$, $35$, and $40$. The marginal constraint on $M_{\mathrm{LMC}}$ improves sharply between $n_{\max}=20$ and $n_{\max}=25$, from $\sigma_{M_{\mathrm{LMC}}} = 1.36 \times 10^{10}~\mathrm{M}_\odot$ to $1.11 \times 10^{10}~\mathrm{M}_\odot$, indicating that radial orders $n \approx 21$--$25$ capture the dominant LMC-sensitive structure at 30--120 kpc.

Beyond $n_{\max} = 30$ the estimated constraints plateau and then loosen; the joint Fisher volume reaches its minimum at $n_{\max} = 30$, $6.0$ per cent tighter than the full $n_{\max} = 40$ expansion. This loosening is finite-sample degradation of the estimated compression, not a loss of physical information. The score covariance of the four summaries stays well conditioned across the sweep (condition number $4.6$--$6.2$), but the radial modes of order $n>30$, though comparable in amplitude to lower orders, have finite-difference derivatives that are markedly less stable across random seeds. These modes let the MOPED projection fit derivative noise and generalise worse to independent validation data.

We use the same $10$ independent, non-fiducial test simulations from the transfer-matrix stability test to check whether $n_{\max}=30$ is specific to the fiducial point. Across these validation runs, truncating at $n_{\max} = 30$ gives $0.7$--$3.1$ per cent tighter marginal constraints than the full $n_{\max} = 40$ expansion ($1.9\%$ for $M_{\mathrm{MW}}$, $1.7\%$ for $M_{\mathrm{LMC}}$, $0.7\%$ for $c$, $3.1\%$ for $q$). This suggests that the degradation from modes beyond the knee is not confined to the fiducial simulation.

\subsection{Information retained by BFE+MOPED alone}
\label{sec:information_interpretability}
\label{sec:fisher_comparison}

We now compare the BFE+MOPED summaries alone against the field-level CFM and velocity-moment benchmarks. The local Fisher matrix of $\mathbf{m}$ is computed at $\boldsymbol{\theta}_\star$ exactly as for the velocity moments (equation~\ref{eq:fisher_moments}), with the mean and $4\times 4$ covariance of $\mathbf{m}$ estimated from the fiducial realisations and the mean derivatives evaluated by finite differences using the derivative simulations. For the BFE+MOPED summaries, however, the Gaussian-likelihood assumption is empirical rather than guaranteed: although the BFE coefficients are particle sums, the radial basis functions contain nodes and steep gradients, so finite-particle coefficients can carry high-leverage contributions and non-Gaussian tails. The Fisher matrix therefore provides only a local forecast, whose likelihood assumptions must be checked with validation diagnostics. Table~\ref{tab:moped_constraint_comparison} compares these Fisher constraints directly with the marginal posteriors from the full phase-space CFM and the velocity-moment Fisher forecast already reported in Table~\ref{tab:cfm_velocity_constraint_comparison}.

The BFE+MOPED constraints lie between the two benchmarks: they are $(3.1,\,1.8,\,3.3,\,1.2)$ times tighter than the velocity-moment forecast for $(M_{\mathrm{MW}}, M_{\mathrm{LMC}}, c, q)$, while remaining $(3.2,\,3.1,\,1.3,\,2.0)$ times looser than the full-phase-space CFM benchmark. Thus the compression recovers part of the field-level gain in summaries whose physical content we examine in Section~\ref{sec:interpretability}.

\begin{table*}
\centering
\caption{Fiducial marginal $1\sigma$ parameter constraints from the field-level CFM benchmark, standard velocity moments, and BFE+MOPED summaries, with the joint BFE+MOPED plus velocity moment combination.}
\label{tab:moped_constraint_comparison}
\begin{tabular}{lcccc}
\hline
Parameter & CFM (phase-space limit) & Velocity moments & BFE+MOPED alone & Joint (19D) \\
\hline
$M_{\mathrm{MW}}$  & 0.066 & 0.655 & 0.212 & 0.188 \\
$M_{\mathrm{LMC}}$ & 0.36  & 2.02  & 1.10  & 1.03 \\
$c$                & 0.52  & 2.26  & 0.676  & 0.651 \\
$q$                & 0.0084& 0.021 & 0.0173 & 0.0147 \\
\hline
\end{tabular}
\par
\medskip
\begin{flushleft}
{\small \textbf{Notes.} Mass constraints are in units of $10^{11}\,\mathrm{M}_\odot$ for $M_{\mathrm{MW}}$ and $10^{10}\,\mathrm{M}_\odot$ for $M_{\mathrm{LMC}}$; $c$ and $q$ are dimensionless. The CFM values are marginal posterior widths from the particle-level likelihood grid. The velocity-moment and BFE+MOPED values are local Fisher forecasts at the same fiducial point. The joint (19D) values are the local Fisher forecast for the concatenated BFE+MOPED plus velocity moment vector measured on common realisations.}
\end{flushleft}
\end{table*}

The gain over velocity moments follows from the angular structure of the LMC perturbation: the reflex response and dynamical-friction wake are anisotropic and spatially localised, so the BFE spherical harmonics capture them where all-sky velocity moments in broad radial shells partly smooth them out. To understand why the BFE+MOPED constraints nonetheless fall short of the CFM benchmark, we test two effects internal to the summary pipeline. First, radial truncation does not remove useful modes: the $n_{\max}$ sweep of Section~\ref{sec:MOPED} shows no improvement beyond $n_{\max}=30$. Second, to test whether parameter-dependent covariance contributes appreciable information, we decompose the Gaussian Fisher matrix into a mean-response term and a covariance-derivative term,
\begin{equation}
F_{\alpha\beta} = \left(\frac{\partial \boldsymbol{\mu}}{\partial \theta_\alpha}\right)^{\!\top} \mathbf{C}^{-1}\, \frac{\partial \boldsymbol{\mu}}{\partial \theta_\beta} + \frac{1}{2}\,\mathrm{Tr}\!\left[\mathbf{C}^{-1}\, \frac{\partial \mathbf{C}}{\partial \theta_\alpha}\, \mathbf{C}^{-1}\, \frac{\partial \mathbf{C}}{\partial \theta_\beta}\right],
\label{eq:fisher_decomposition}
\end{equation}
where $\boldsymbol{\mu}$ and $\mathbf{C}$ are the summary mean and covariance. The covariance term is negligible, at $\sim5\times10^{-4}$ of the mean-response term and changing the Fisher volume by only $0.09$ per cent. The residual gap to CFM is therefore not driven by radial-order truncation or by parameter-dependent covariance within the BFE+MOPED summaries, but by information outside the four linear projections, including the higher-order and non-Gaussian phase-space correlations available to the particle-level neural transport likelihood.

\subsection{Physical interpretability of BFE+MOPED channels}
\label{sec:interpretability}

The Fisher tests quantify how much information the BFE+MOPED summaries retain. We now identify which phase-space structures supply that information. Although the MOPED projection preserves the Fisher information of the BFE features, the resulting weight vectors are linear combinations of many spatial and kinematic modes. We therefore back-project the MOPED weights to all-sky maps of the per-particle contributions to each summary. This per-particle decomposition is possible because both the BFE projection and the MOPED transformation are linear.

In practice, we draw the particles from the fiducial simulation in the $60$--$90$~kpc Galactocentric shell, calculate the single-particle BFE features, and apply the fiducial MOPED transfer matrix channel by channel to obtain the per-particle contribution $q^{(i)}_{\alpha,c}$ to the $\alpha$-th compressed summary from each BFE channel $c$.

For the first-order velocity channels, the three Cartesian momentum contributions $(b_x, b_y, b_z)$ are rotated into the local Galactocentric spherical basis to give the radial, latitudinal, and longitudinal components $(v_r, v_b, v_\ell)$. The resulting per-particle scores are smoothed onto a Mollweide grid in Galactocentric longitude and latitude for visualisation.

\begin{figure*}
	\includegraphics[width=\textwidth]{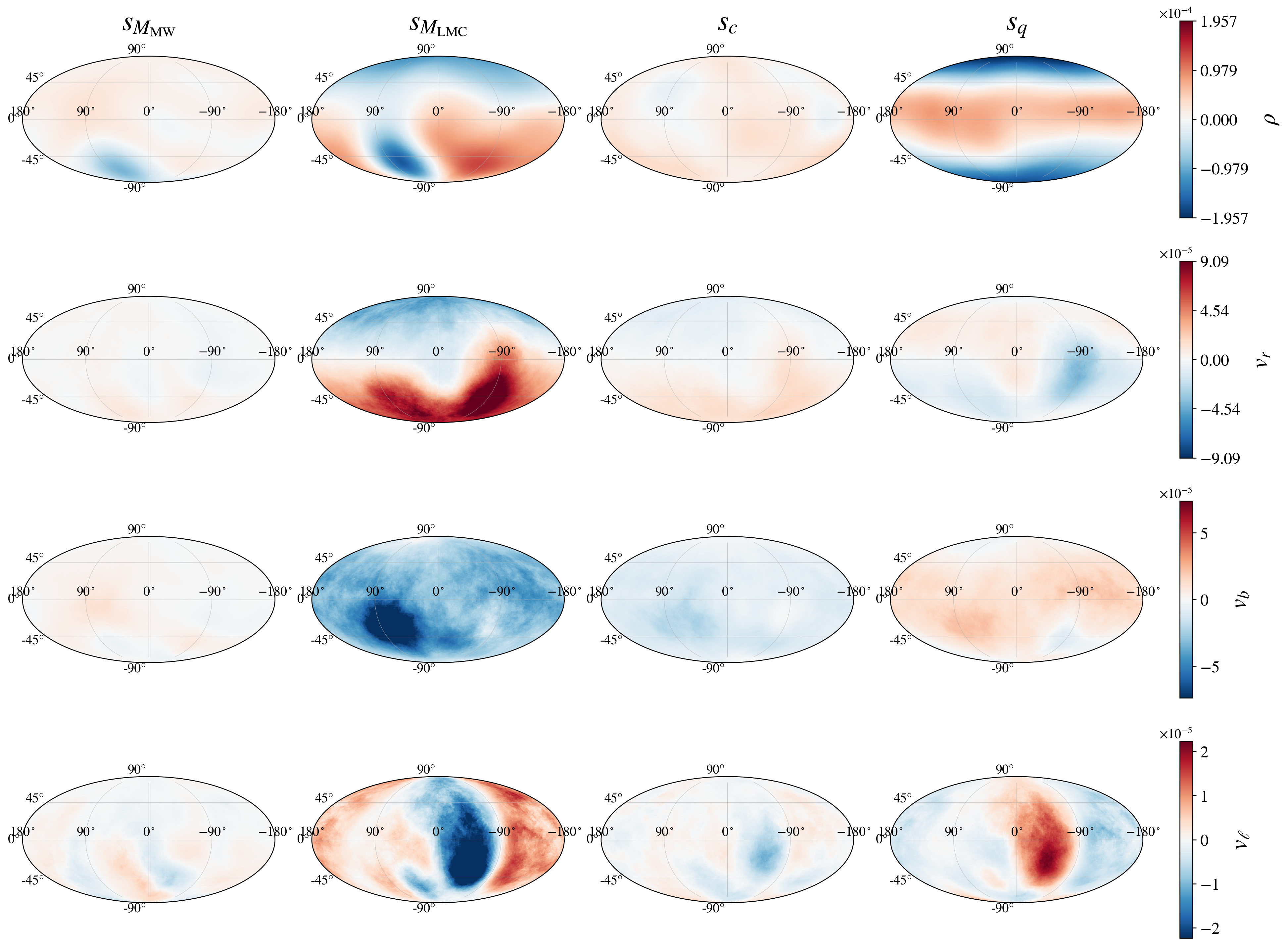}
    \caption{Per-particle MOPED contribution maps from the density and first-moment velocity BFE channels in the $60$--$90$~kpc shell. Columns correspond to the four scalar BFE+MOPED summaries, and rows correspond to $\rho$, $v_r$, $v_b$, and $v_\ell$. These maps are not maps of the fields themselves; they show where particles in each channel increase or decrease each summary. The sign is conventional, while the angular pattern identifies the structures selected by MOPED.}
    \label{fig:moped_first_order}
\end{figure*}

Figure~\ref{fig:moped_first_order} shows how the density and first-moment velocity channels enter the four scalar summaries. Each column is a parameter summary and each row a physical channel, so a bright region marks where particles in that channel and sky direction contribute strongly to the summary, not where the underlying field is largest. Red and blue give the sign of the MOPED weight, which is arbitrary under the QR convention, so the interpretable content is the angular morphology and the physical channel.

The first-order maps show that $s_{M_{\mathrm{LMC}}}$ is tied to the LMC-induced asymmetry. Its density and velocity panels select the same large-scale reflex dipole and local wake seen in the BFE reconstruction in Figure~\ref{fig:bfe_mollweide}. The strongest radial-velocity contribution aligns with the wake region, while the tangential-velocity panels capture coherent large-scale reflex motion signatures, including a global bias in latitudinal velocity ($v_{b}$). Thus the LMC summary is not just a high-dimensional coefficient combination. It can be traced back to the expected reflex and wake response.

The same figure shows a different morphology for $s_q$. The density panel is dominated by a quadrupolar pattern, with opposite contributions near the equatorial and polar regions. This is the expected angular signature of changing the halo flattening at fixed radius. The first-order velocity panels carry weaker but coherent structures. The $s_{M_{\mathrm{MW}}}$ and $s_c$ columns show little structure, for different reasons: the MW-mass response is dominated by the velocity-dispersion channels discussed below, while the largest single contribution to $s_c$ is the angularly uniform $\ell=0$ density monopole ($24\%$; Table~\ref{tab:moped_signal_decomposition}), which produces no visible all-sky morphology.

To complete the channel interpretation, we also back-project the diagonal second-moment velocity channels. The six independent second-moment contributions $c_{kl}$ are contracted with the local $(\hat{\mathbf{e}}_r, \hat{\mathbf{e}}_b, \hat{\mathbf{e}}_\ell)$ basis to give the diagonal velocity-dispersion contributions $(v_r^{2}, v_b^{2}, v_\ell^{2})$ in the same $60$--$90$~kpc shell.

\begin{figure*}
	\includegraphics[width=\textwidth]{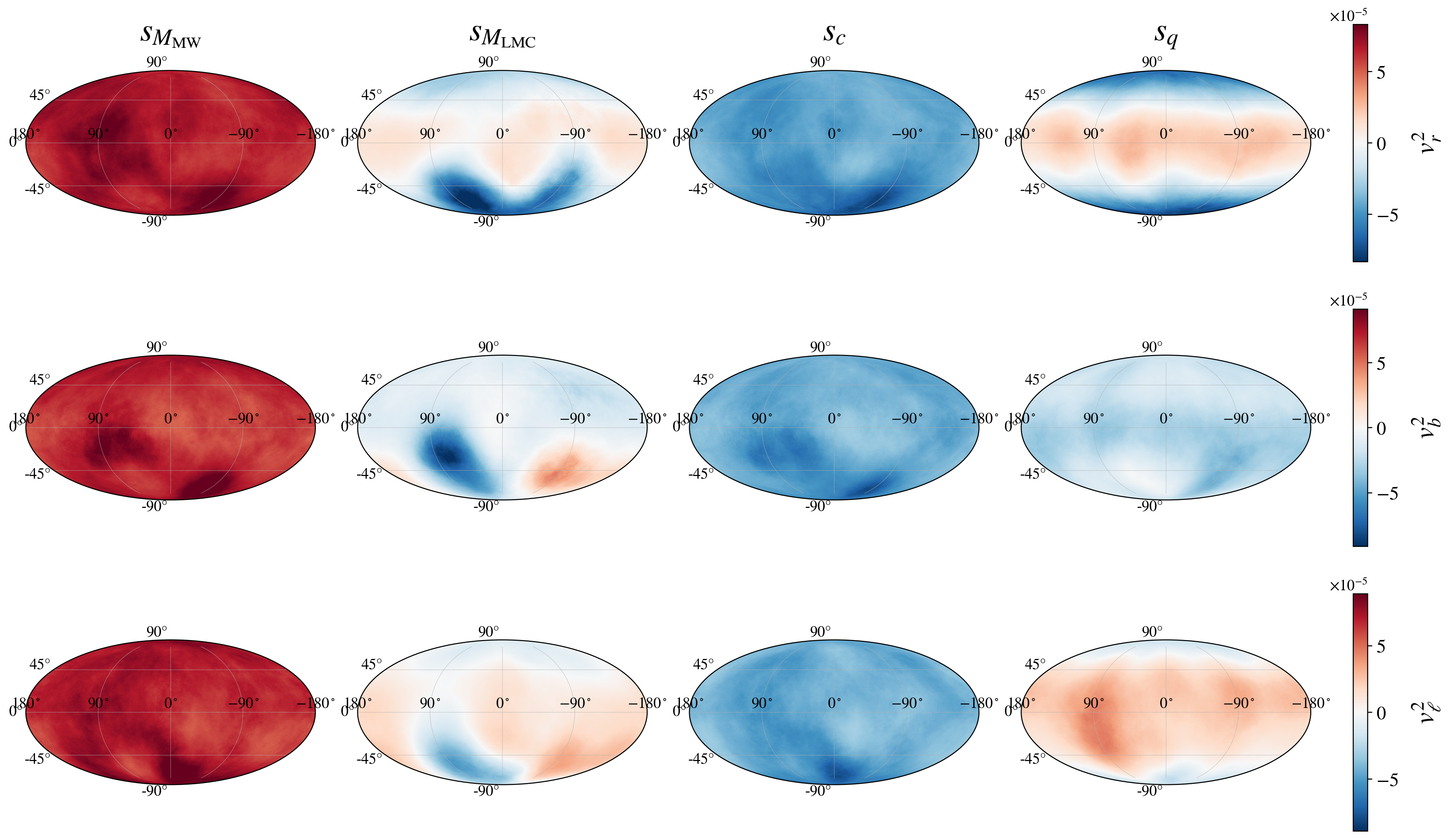}
    \caption{Per-particle MOPED contribution maps from the diagonal second-moment velocity channels in the $60$--$90$~kpc shell. These maps show where $v_r^2$, $v_b^2$, and $v_\ell^2$ contribute to each scalar summary. The dominant response is in $s_{M_{\mathrm{MW}}}$ and $s_c$, whose nearly monopolar patterns reflect the MW mass--concentration degeneracy in tracer dispersions.}
    \label{fig:moped_second_order}
\end{figure*}

Figure~\ref{fig:moped_second_order} completes the interpretation by showing the velocity-dispersion channels. The $s_{M_{\mathrm{MW}}}$ maps are nearly monopolar in all three $v^{2}$ rows, because increasing the MW mass raises the all-sky velocity dispersion in the shell; the $s_c$ maps show a weaker monopolar response of opposite sign, a secondary contribution that complements the density monopole and first-moment terms discussed above. Since increasing either $M_{\mathrm{MW}}$ or $c$ raises the enclosed mass in the shell and therefore the dispersion, MOPED assigns opposite weights to this shared response to separate an overall mass increase from a change in radial profile shape. This is the velocity-space form of the NFW $M_{200}$--$c_{200}$ degeneracy in tracer kinematics \citep[e.g.,][]{Cautun2020}. The weaker structures in the $s_{M_{\mathrm{LMC}}}$ and $s_q$ columns are consistent with the dipolar and quadrupolar signatures seen in the first-order channels.

We quantify the same back-projection by grouping the per-particle MOPED weights into physical channels. For each MOPED summary $s_\alpha$ and channel group $G$, we sum the absolute values of the single-particle contributions $q^{(i)}_{\alpha,c}$ over all particles $i$ in the $60$--$90$~kpc shell and all BFE channels $c$ assigned to that group:
\begin{equation}
P_{\alpha,G} =
\frac{\sum_{c\in G}\sum_i |q^{(i)}_{\alpha,c}|}
{\sum_{G'}\sum_{c\in G'}\sum_i |q^{(i)}_{\alpha,c}|}.
\label{eq:channel_decomposition}
\end{equation}
The density coefficients are grouped into $\ell=0$, $\ell=1$, and $\ell\geq2$ multipoles. The first-moment velocity coefficients are grouped into $(v_r,v_b,v_\ell)$, and the diagonal second-moment coefficients are grouped into $(v_r^2,v_b^2,v_\ell^2)$. The absolute value is used because the MOPED weight signs are conventional and opposite-sign sky regions can otherwise cancel; the percentages therefore measure which physical channels supply the amplitude of each summary in this shell, as a local diagnostic tied to the fiducial transfer matrix.

\begin{table*}
\centering
\caption{Channel-wise signal decomposition of the BFE+MOPED summaries in the $60$--$90$~kpc shell. This table is an interpretability diagnostic for $\mathbf{m}$, not the mutual-information complementarity test.}
\label{tab:moped_signal_decomposition}
\definecolor{brandblue}{rgb}{0.15, 0.35, 0.65}
\begin{tabular}{lccccccccc}
\hline
& \multicolumn{3}{c}{Density multipoles $\rho$} & \multicolumn{3}{c}{First-moment velocities $v$} & \multicolumn{3}{c}{Second-moment velocities $v^2$} \\
MOPED Summary & $\ell=0$ & $\ell=1$ & $\ell\ge 2$ & $v_r$ & $v_b$ & $v_\ell$ & $v_r^2$ & $v_b^2$ & $v_\ell^2$ \\
\hline
$s_{M_{\mathrm{MW}}}$  & 4\% & 5\% & 3\% & 4\% & 3\% & 3\% & \cellcolor{brandblue!30}\textbf{25\%} & \cellcolor{brandblue!30}\textbf{26\%} & \cellcolor{brandblue!30}\textbf{26\%} \\
$s_{M_{\mathrm{LMC}}}$ & 7\% & 8\% & 5\% & \cellcolor{brandblue!30}\textbf{23\%} & \cellcolor{brandblue!30}\textbf{27\%} & \cellcolor{brandblue!15}\textbf{10\%} & 8\% & 8\% & 4\% \\
$s_c$                  & \cellcolor{brandblue!30}\textbf{24\%} & \cellcolor{brandblue!15}\textbf{10\%} & \cellcolor{brandblue!15}\textbf{11\%} & 6\% & 7\% & 4\% & \cellcolor{brandblue!15}\textbf{12\%} & \cellcolor{brandblue!15}\textbf{13\%} & \cellcolor{brandblue!15}\textbf{12\%} \\
$s_q$                  & \cellcolor{brandblue!15}\textbf{14\%} & \cellcolor{brandblue!15}\textbf{16\%} & 7\% & \cellcolor{brandblue!15}\textbf{11\%} & \cellcolor{brandblue!15}\textbf{13\%} & 9\% & \cellcolor{brandblue!15}\textbf{11\%} & \cellcolor{brandblue!15}\textbf{10\%} & 9\% \\
\hline
\end{tabular}
\par
\medskip
\begin{flushleft}
{\small \textbf{Notes.} The percentages are the channel fractions $P_{\alpha,G}$ defined in equation~\ref{eq:channel_decomposition}. Cell backgrounds highlight dominant channels. High signal shares ($\ge 20\%$) are shaded in dark blue, and moderate shares ($10\%$--$19\%$) are shaded in light blue. The entries describe which BFE channels contribute to each BFE+MOPED summary.}
\end{flushleft}
\end{table*}

Table~\ref{tab:moped_signal_decomposition} quantifies the same pattern: $s_{M_{\mathrm{LMC}}}$ is dominated by the first-moment velocity channels ($60\%$) plus the $\ell=1$ density dipole ($8\%$), $s_{M_{\mathrm{MW}}}$ by the second-moment velocity channels ($77\%$), and $s_c$ splits between the density monopole ($24\%$, its largest single channel) and the dispersion channels ($37\%$ in total), while $s_q$ draws non-negligible signal from all three sectors. These fractions explain the physical content of the BFE+MOPED summaries, but they do not by themselves show which information is missing relative to the standard velocity moments or the field-level benchmark.

\section{Complementarity of BFE and velocity moments}
\label{sec:joint_constraints}

BFE+MOPED summaries and velocity moments are two summaries of the same tracer catalogue, and they need not be treated as mutually exclusive alternatives. The BFE+MOPED summaries retain angular structure in the density and velocity fields, including the reflex dipole, wake, and halo-shape response, while the velocity moments provide a compact measurement of the all-sky mean and dispersion profiles in radial shells. Since these summaries compress different projections of the same phase-space perturbation, the relevant question is not only which one performs better in isolation, but whether their combination retains information that either summary loses alone. This section first uses mutual information to test whether the two summaries are complementary, and then reports the corresponding joint Fisher constraints.

\subsection{Complementarity from Mutual Information}
\label{sec:mutual_information}

Fisher constraints rank summaries by the marginal $1\sigma$ constraints they provide on each parameter, assuming a Gaussian summary likelihood and a local comparison around the fiducial point. To go beyond these assumptions, we compare the summary representations using their total information content. Let $\mathbf{s}$ denote any summary vector derived from the same tracer catalogue, with $\mathbf{m}$ denoting the four BFE+MOPED summaries and $\mathbf{v}$ denoting the 15 standard velocity moments. The mutual information (MI) between the physical parameters $\boldsymbol{\theta}$ and a summary vector $\mathbf{s}$ is
\begin{equation}
I(\boldsymbol{\theta}; \mathbf{s}) = \mathbb{E}_{p(\boldsymbol{\theta}, \mathbf{s})}\!\left[\log \frac{p(\boldsymbol{\theta}, \mathbf{s})}{p(\boldsymbol{\theta})\,p(\mathbf{s})}\right].
\label{eq:mi_definition}
\end{equation}
MI is parameterisation-invariant and assumes neither a Gaussian likelihood nor a local expansion around a fiducial point. It measures how much a summary contracts the posterior volume on average over the prior, so a higher MI means the summary retains more of the information in the raw 6D phase-space data. The joint mutual information $I(\boldsymbol{\theta}; \mathbf{m}, \mathbf{v})$ of the combined BFE+MOPED and velocity-moment summaries can be decomposed using the chain rule:
\begin{equation}
I(\boldsymbol{\theta}; \mathbf{m}, \mathbf{v}) = I(\boldsymbol{\theta}; \mathbf{m}) + I(\boldsymbol{\theta}; \mathbf{v}\mid \mathbf{m}),
\label{eq:mi_chain_rule}
\end{equation}
where the conditional mutual information $I(\boldsymbol{\theta}; \mathbf{v}\mid \mathbf{m}) = I(\boldsymbol{\theta}; \mathbf{m}, \mathbf{v}) - I(\boldsymbol{\theta}; \mathbf{m})$ quantifies the additional information contributed by the velocity moments after the BFE+MOPED summaries are already known. A non-zero conditional MI indicates that the two summary sets retain complementary information from the same phase-space dataset.

We estimate the mutual information for the three summary representations using a variational Barber--Agakov lower bound, modeling the variational posterior with conditional flow matching. The detailed neural architecture, flow-matching training procedure, and backward ODE integration settings are described in Appendix~\ref{app:mi_details}. Table~\ref{tab:mi_comparison} reports the resulting mutual information estimates and the corresponding expected posterior volume contraction factors $\exp(I)$.

\begin{table*}
\centering
\caption{Mutual information and expected posterior volume contraction factors for the different summary representations. This table is the diagnostic of complementarity between BFE+MOPED summaries and velocity moments.}
\label{tab:mi_comparison}
\begin{tabular}{lcc}
\hline
Summary representation & Mutual Information $I(\boldsymbol{\theta}; \mathbf{s})$ [nats] & Volume contraction $\exp(I)$ \\
\hline
Velocity moments $\mathbf{v}$ (15 dims) & 7.07 & $1{,}176$ \\
BFE+MOPED $\mathbf{m}$ (4 dims) & 7.72 & $2{,}253$ \\
Joint summaries $(\mathbf{m}, \mathbf{v})$ (19 dims) & 8.46 & $4{,}722$ \\
\hline
\multicolumn{3}{l}{\textbf{Information decomposition:}} \\
Shared information $I(\boldsymbol{\theta}; \mathbf{m} \cap \mathbf{v})$ & 6.33 & $561$ \\
Conditional moment info $I(\boldsymbol{\theta}; \mathbf{v} \mid \mathbf{m})$ & 0.74 & $2.1$ \\
Conditional MOPED info $I(\boldsymbol{\theta}; \mathbf{m} \mid \mathbf{v})$ & 1.39 & $4.0$ \\
\hline
\end{tabular}
\par
\medskip
\begin{flushleft}
{\small \textbf{Notes.} The expected posterior volume contraction factor relative to the prior is given by $\exp(I)$, where $I$ is the mutual information in nats. The shared information is defined as $I(\boldsymbol{\theta};\mathbf{m}) + I(\boldsymbol{\theta};\mathbf{v}) - I(\boldsymbol{\theta};\mathbf{m},\mathbf{v})$. The conditional rows quantify the information added by one summary set after the other is known.}
\end{flushleft}
\end{table*}

Table~\ref{tab:mi_comparison} quantifies the complementarity. The joint summary vector gives a larger expected posterior-volume contraction than either summary alone, and the conditional rows give the added value of one summary once the other is known: velocity moments add a factor of $2.1$ beyond BFE+MOPED, while BFE+MOPED adds a factor of $4.0$ beyond velocity moments. Since both factors exceed unity, combining the summaries is worthwhile; the joint Fisher forecast in Section~\ref{sec:joint_fisher} quantifies that gain locally.

\subsection{Joint Fisher forecast setup}
\label{sec:joint_fisher}

The mutual-information analysis motivates a local Fisher test of the combined summary vector. We form the joint vector by directly concatenating the four BFE+MOPED summaries and the 15 standard velocity moments,
\begin{equation}
\mathbf{y} = (\mathbf{m}, \mathbf{v}) \in \mathbb{R}^{19}.
\end{equation}
The joint Fisher matrix is then computed from the covariance and parameter-response matrix of this full 19-dimensional vector at the fiducial point. Because $\mathbf{m}$ and $\mathbf{v}$ are measured from the same particle catalogue, the full covariance retains their cross-correlations (median absolute value $0.05$, maximum $0.42$), rather than treating the two Fisher matrices as additive.

The resulting joint constraints, listed in the final column of Table~\ref{tab:moped_constraint_comparison}, are marginal $1\sigma$ values of $0.188\times10^{11}\,\mathrm{M}_\odot$, $1.03\times10^{10}\,\mathrm{M}_\odot$, $0.651$, and $0.0147$ for $(M_{\mathrm{MW}}, M_{\mathrm{LMC}}, c, q)$. These tighten the BFE+MOPED-only constraints by $11.4$, $7.2$, $3.7$, and $15.1$ per cent and the velocity-moment constraints by $71.1$, $49.1$, $71.0$, and $30.0$ per cent. Relative to the field-level benchmark, the joint vector closes $16.6$, $10.6$, $16.1$, and $29.5$ per cent of the BFE+MOPED-to-CFM gap. Measured from the velocity-moment baseline instead, the joint summaries recover $50$--$92$ per cent of the constraint-width improvement achieved by the field-level likelihood, and the remaining factor of $1.3$--$2.9$ quantifies the information that the current linear compression does not capture.

\begin{figure}
    \includegraphics[width=\columnwidth]{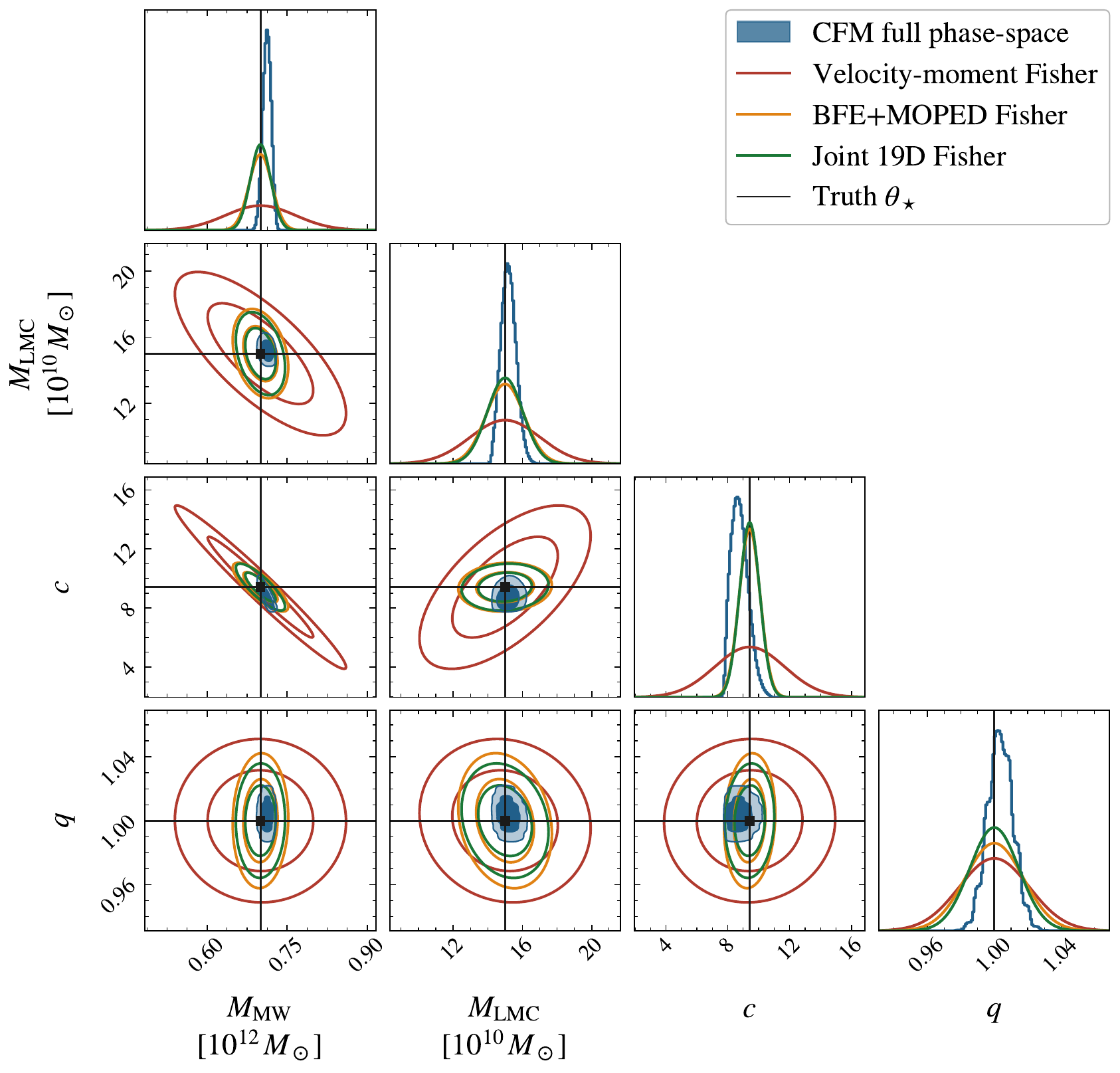}
    \caption{Fiducial marginal constraints from all summary sets against the field-level CFM benchmark, shown as $68$ and $95$ per cent credible regions in the same convention as Figure~\ref{fig:cfm_fiducial_benchmark}. The CFM full phase-space posterior (blue, filled) is overlaid with the velocity-moment (red), BFE+MOPED (orange), and joint 19-dimensional (green) Fisher forecasts. The forecasts nest from the loosest velocity moments through BFE+MOPED to the tightest joint vector, which still does not reach the CFM posterior.}
    \label{fig:all_summaries_benchmark}
\end{figure}

Figure~\ref{fig:all_summaries_benchmark} places all four cases in the same parameter planes and $68$ and $95$ per cent convention as Figure~\ref{fig:cfm_fiducial_benchmark}. The Fisher contours nest from the velocity moments through BFE+MOPED to the joint vector, which lies closest to the CFM posterior but still encloses a visibly larger region. Consistent with Table~\ref{tab:mi_comparison}, the joint contours sit only slightly inside the BFE+MOPED ones but far inside the velocity-moment ones: the $2.1$-fold volume contraction from adding velocity moments is shared across all four parameters, so it appears as the modest $3.7$--$15.1$ per cent marginal tightenings reported above, whereas adding BFE+MOPED to the moments yields the larger $4.0$-fold gain.

\section{Simulation-Based Inference on summary statistics}
\label{sec:implementation}

This section describes how the density and velocity summaries are converted into constraints on the MW--LMC parameters. We define the summary likelihood, build a likelihood emulator to interpolate to off-grid parameter, test its calibration on held-out HaloDance simulations, and measure its sensitivity to radially biased tracers. These tests establish the accuracy and current limitations of the summary-level inference before application to observational data.

The summary-likelihood emulator is used for two closely related inference runs. The primary summary-level constraints use the joint vector $\mathbf{y}=(\mathbf{m},\mathbf{v})\in\mathbb{R}^{19}$, where $\mathbf{m}$ contains the four BFE+MOPED summaries and $\mathbf{v}$ contains the 15 velocity moments. The same emulator architecture is also applied to the four-dimensional BFE+MOPED vector $\mathbf{m}$ for the tracer-anisotropy stress test of Section~\ref{sec:summary_validation}. In both cases the catalogue is mapped to a summary vector using a fixed MOPED compression operator calibrated at the $\mathbf{\theta_{\star}}$, and the emulator provides the likelihood of that summary as a function of the HaloDance parameters.

We first check that a Gaussian likelihood is adequate for the primary 19-dimensional joint summaries. Pooling $500$ realisations at each of $185$ held-out parameter points, the whitened Mahalanobis $d^{2}$ closely follows the expected $\chi^2_{19}$ distribution (mean $18.96$ and variance $36.4$ against the expected $19$ and $38$; empirical $68$ and $95$ per cent coverages of $68.1$ and $95.4$ per cent), and the marginal skewness and excess kurtosis never exceed $0.026$ and $0.040$ in absolute value. The four-dimensional BFE+MOPED summaries pass the same diagnostic and are used for the anisotropy test below.

Let $\mathbf{s}$ denote the summary vector used in a given run, with $\mathbf{s}=\mathbf{y}$ for the primary joint analysis and $\mathbf{s}=\mathbf{m}$ for the BFE+MOPED-only stress test. We model the summary distribution at fixed parameters as
\begin{equation}
\mathbf{s} \mid \boldsymbol{\theta} \sim \mathcal{N}\!\bigl(\boldsymbol{\mu}(\boldsymbol{\theta}), \boldsymbol{\Sigma}(\boldsymbol{\theta})\bigr),
\end{equation}
and emulate the mean and covariance with a mixture-density-network-style (MDN-style) Gaussian likelihood emulator. In the present implementation the mixture has a single multivariate Gaussian component, so the network predicts the conditional mean and covariance of $p(\mathbf{s}\mid\boldsymbol{\theta})$ rather than a flow or a particle-level density. For each training parameter point, we draw 500 independent realisations of 5,000 particles from the 30--120~kpc shell and pass each through the chosen summary compression. These realisations define an empirical mean $\hat{\boldsymbol{\mu}}(\boldsymbol{\theta})$ and covariance $\hat{\boldsymbol{\Sigma}}(\boldsymbol{\theta})$.

An ensemble of five independently trained networks accounts for variation between training runs. The effective likelihood uses the ensemble-averaged mean $\boldsymbol{\mu}_{\rm eff}(\boldsymbol{\theta})$ and a covariance $\boldsymbol{\Sigma}_{\rm eff}(\boldsymbol{\theta})$ that includes both the predicted summary covariance and the scatter among network mean predictions. Appendix~\ref{app:sbi_details} gives details of the network architecture, training objective and covariance construction.

Given an observed summary vector $\mathbf{s}_{\rm obs}$, the emulator likelihood is
\begin{equation}
\log \mathcal{L}(\boldsymbol{\theta}) = -\frac{1}{2}\left[\boldsymbol{\delta}^{\top}\boldsymbol{\Sigma}_{\rm eff}^{-1}\boldsymbol{\delta}+\log\det\boldsymbol{\Sigma}_{\rm eff}\right],
\end{equation}
where $\boldsymbol{\delta}=\mathbf{s}_{\rm obs}-\boldsymbol{\mu}_{\rm eff}(\boldsymbol{\theta})$. The prior is uniform over the HaloDance parameter box and zero outside it. Posterior sampling uses \textsc{dynesty} dynamic nested sampling with multi-ellipsoid bounding and the \texttt{rwalk} sampler, initialized with 500 live points and 200 live points per dynamic batch, an evidence stopping threshold $\Delta\ln\mathcal{Z}_{\rm init}=0.05$, and a target posterior effective sample size of 4,000.

This summary-level MDN-style emulator is distinct from the particle-level CFM in Section~\ref{sec:cfm}. The particle-level CFM operates on individual 6D phase-space coordinates and defines the field-level information benchmark, whereas the summary-level emulator operates on $\mathbf{s}$ and provides the practical posterior sampler over the HaloDance parameter space. This separation is useful for planned H3 and SDSS-V applications, because the observed catalogue only has to be converted once into the adopted summary vector. Survey masks, tracer selection, and anisotropy priors can then be tested at the summary-likelihood level.

\subsection{Summary-level validation and systematics}
\label{sec:summary_validation}

\begin{figure*}
	\includegraphics[width=\textwidth]{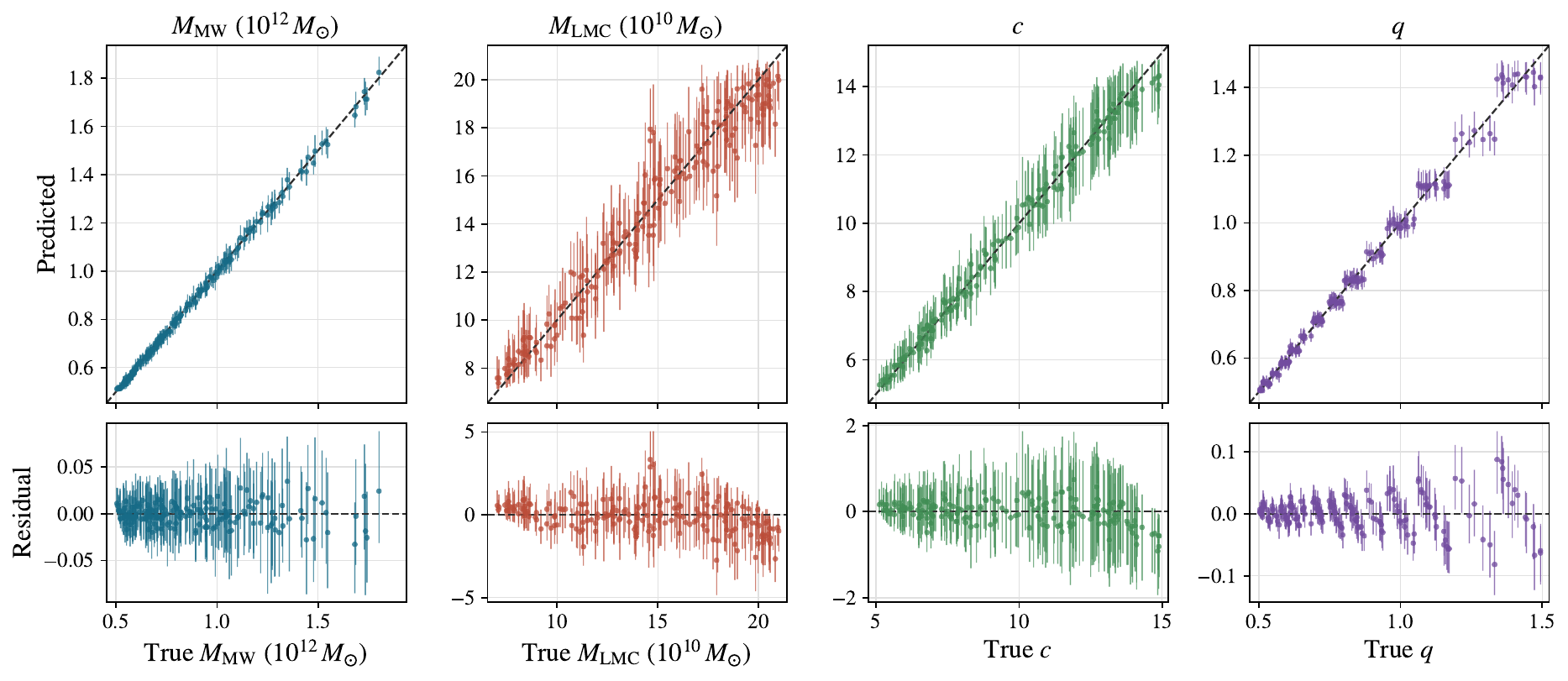}
    \caption{True-vs-predicted accuracy of the MDN-style Gaussian summary-likelihood emulator for the primary joint-summary pipeline on the isotropic 185-grid validation set. The top row shows posterior medians against the true parameters, and the bottom row shows median-minus-true residuals.}
    \label{fig:true_vs_predicted}
\end{figure*}

The primary summary-level SBI pipeline is trained and validated on HaloDance halos with an isotropic velocity distribution ($\beta=0$), using the joint $19$-dimensional summary vector $\mathbf{y}=(\mathbf{m},\mathbf{v})$. Figure~\ref{fig:true_vs_predicted} shows the posterior medians and their residuals on the held-out validation set. Appendix~\ref{app:sbi_validation} provides the corresponding quantitative residual measurements, marginal P-P ranks, and joint TARP coverage. Because the velocity moments intentionally include dispersion information that responds directly to tracer anisotropy, the robustness experiment below uses a separately trained four-dimensional BFE+MOPED emulator to isolate the response of the BFE density and velocity channels.

The MDN-style likelihood remains an approximation because it uses a single multivariate Gaussian component. Section~\ref{sec:implementation} and Appendix~\ref{app:sbi_validation} show that this approximation is adequate for the primary joint summaries in the isotropic validation set, and the four-dimensional BFE+MOPED summaries pass the same fixed-parameter Gaussianity check for the stress test. For survey applications, masks, selection functions, and tracer-population mixtures may make the whitened summaries less Gaussian. In that case, the same summary interface can be retained while replacing the single-Gaussian likelihood with a normalising flow or a summary-level CFM model for $p(\mathbf{s}\mid\boldsymbol{\theta})$.

Observed outer-halo tracers are radially biased \citep[e.g.,][]{Chandra2025,Li2026b}, so tracer anisotropy remains a leading systematic for applying these summaries to data. To test this mismatch, we apply the isotropic-trained BFE+MOPED emulator to the separate radially varying HaloDance set of Section~\ref{sec:halodance}, with $\beta(r) = -0.15 - 0.2\,\alpha(r)$, corresponding to $\beta(r)\simeq0.3$--$0.45$ across $30$--$120$~kpc for the halos used here. This is a stress test of model misspecification, not a retrained anisotropic inference model. We select 148 non-training parameter points, draw 10 independent 5,000-particle realisations per point, and process each realisation with the same fiducial BFE+MOPED transfer matrix, MDN-style Gaussian emulator, and \textsc{dynesty} sampler used for the isotropic BFE+MOPED validation.

\begin{figure*}
	\includegraphics[width=\textwidth]{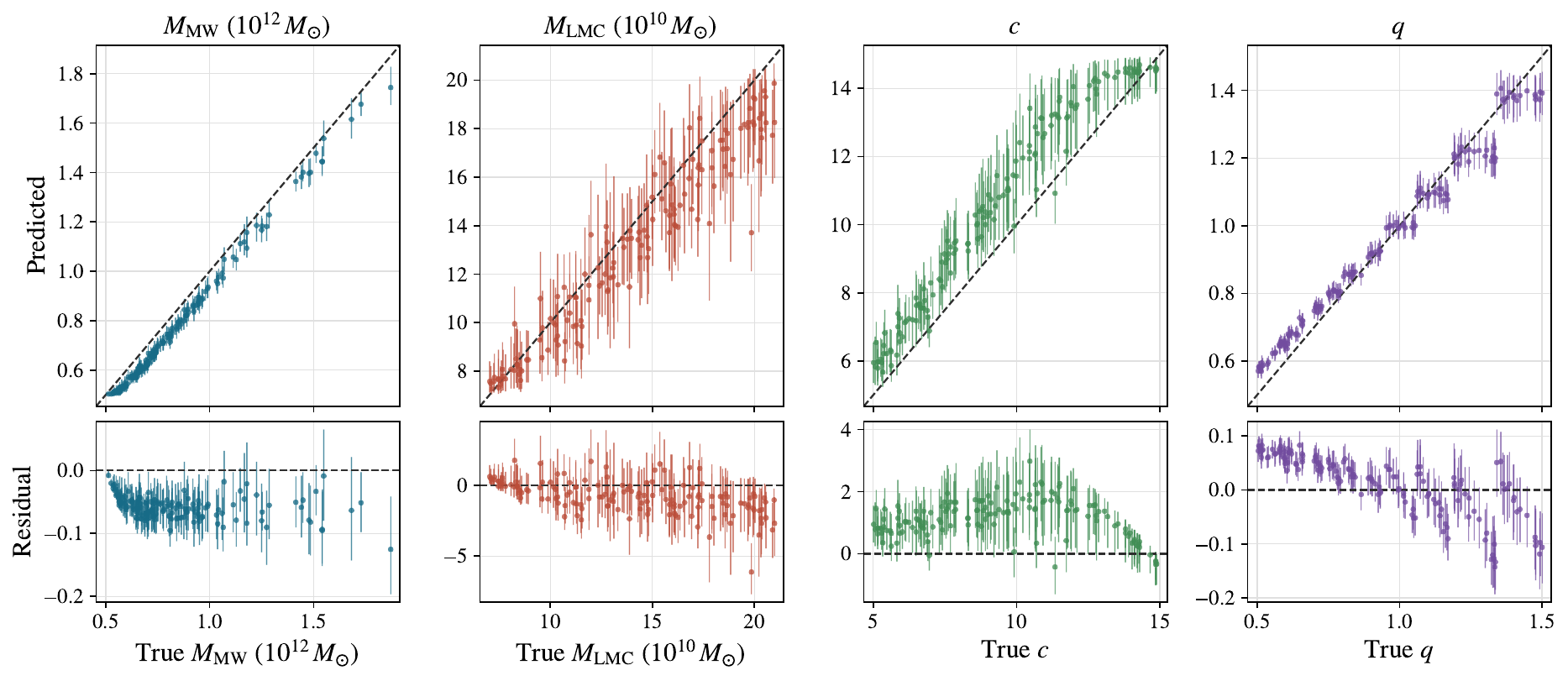}
    \caption{True-vs-predicted accuracy for the anisotropic stress test of the isotropic-trained BFE+MOPED-only pipeline, using the radially varying $\beta(r)$ HaloDance set. Relative to the corresponding isotropic BFE+MOPED validation, the medians shift toward higher $c$ and lower $M_{\mathrm{MW}}$, while the shifts in $M_{\mathrm{LMC}}$ and $q$ are smaller.}
    \label{fig:true_vs_predicted_anisotropic}
\end{figure*}

Figure~\ref{fig:true_vs_predicted_anisotropic} shows that the posterior widths change little in this anisotropic stress test, but the posterior medians shift. The largest shifts are a high bias in concentration, $\Delta c \simeq +1.24$, and a low bias in MW mass, $\Delta M_{\mathrm{MW}} \simeq -0.59 \times 10^{11}\,\mathrm{M}_\odot$. The shifts in $M_{\mathrm{LMC}}$ and $q$ are smaller, with $\Delta M_{\mathrm{LMC}} \simeq -0.82\times 10^{10}\,\mathrm{M}_\odot$ and $\Delta q \simeq +0.024$. The per-grid 68\% intervals contain the truth in only $6.8\%$, $80.4\%$, $19.6\%$, and $40.5\%$ of the anisotropic grids for $(M_{\mathrm{MW}},M_{\mathrm{LMC}},c,q)$, compared with $99.5\%$, $88.7\%$, $97.8\%$, and $84.9\%$ in the isotropic case.

The channel decomposition in Table~\ref{tab:moped_signal_decomposition} explains the direction of the failure. The $M_{\mathrm{LMC}}$ summary draws most of its signal from first-moment velocities and the $\ell=1$ density dipole, which are less sensitive to velocity anisotropy. In contrast, $s_{M_{\mathrm{MW}}}$ receives much of its signal ($77\%$) from the second-moment velocity channels, where radial anisotropy directly changes $\sigma_{v_r}$ at fixed potential, while $s_c$ retains a substantial second-moment component ($37\%$) alongside its leading $\ell=0$ density monopole. An isotropic-trained emulator can therefore absorb anisotropy mismatch as a change in the inferred MW mass and concentration.

The summary-statistic approach also gives a practical way to diagnose and control systematics. Because the summaries can be decomposed by density, first-moment velocity, and second-moment velocity channels, one can test whether a posterior shift is driven by a physically suspect part of the data vector. In a survey application, the same framework can be rerun after censoring selected channels or radial shells, or after adding nuisance parameters for tracer anisotropy, selection, distance errors, and contamination. This does not remove the need for simulation validation, but it makes the failure mode more local than in a less transparent field-level likelihood.

The anisotropy stress test is a first example of this audit. It identifies the second-moment channels as the place where the isotropic training assumption enters most strongly, and it points to the nuisance parameters and channel tests needed before applying the method to survey data.

\section{Discussion}
\label{sec:discussion}

The results define a hierarchy of information content for this simulation setup. The particle-level CFM likelihood gives the tightest local constraints and therefore serves as the field-level reference. Standard velocity moments are more compressed and more interpretable, but lose angular phase-space structure. BFE+MOPED summaries sit between these cases. They recover part of the CFM gain over velocity moments while allowing the contributing density, first-moment velocity, and second-moment velocity channels to be inspected. The mutual-information test then shows that BFE+MOPED and velocity moments are not redundant, motivating the joint summary vector that we adopt for the primary summary-level constraints.

\subsection{Field-level limit and interpretable summaries}

The field-level CFM likelihood is useful here because modern generative density models make a particle-level likelihood for a high-dimensional tracer catalogue computationally tractable. In this work the CFM is not only a sampler. Through the continuous-normalising-flow form of the probability-flow ODE, it gives an evaluable likelihood for each 6D particle conditioned on the MW--LMC parameters. This technical step turns the raw HaloDance catalogue into a field-level information benchmark.

We use flow matching rather than a conventional normalising flow or a score-based diffusion model for this benchmark. Standard normalising flows provide exact likelihoods, but at fixed architecture they can be less flexible for the complex phase-space structure of the perturbed halo. Score-based diffusion models are expressive samplers, but their likelihoods usually require approximate probability-flow integration or variational bounds. CFM keeps the continuous-flow likelihood machinery while using a flexible generative training objective, making it a suitable compromise for a likelihood benchmark.

This benchmark should not be read as an argument for applying field-level inference uncritically to data. Quantifying emulator systematics remains difficult because the likelihood can be precise for the trained simulation distribution while still inheriting simulation mismatch, selection effects, tracer-population mismatch, substructure, velocity-anisotropy variation, and baryonic systematics. The goal of this paper is therefore narrower. The field-level likelihood is used as an information limit, and the summary-level analysis asks how much of that limit can be approached with statistics whose physical content can be inspected.

This framing follows a broader movement in cosmology and field-level inference to evaluate summary statistics by their sufficiency and complementarity rather than by convenience alone \citep[e.g.,][]{Sui2025}. In the MW--LMC problem, the same question has a direct physical interpretation. Velocity moments measure the all-sky radial-bin response of the tracer kinematics and therefore constrain the spherically averaged potential. BFE+MOPED summaries retain angular structure in the density and velocity fields, including the reflex dipole, the local wake, and the halo-shape quadrupole. Their overlap is large because both respond to the same gravitational perturbation, but the conditional mutual information is non-zero because each retains a different projection of that perturbation. Crucially, the four BFE+MOPED summaries are more informative than the fifteen velocity moments despite their lower dimension, giving tighter Fisher constraints on every parameter (factors of $1.2$--$3.3$) and a higher mutual information ($7.72$ against $7.07$ nats). When the signal is anisotropic, retaining its angular structure matters more than the raw number of summaries.

\subsection{Complementarity and missing information}
\label{sec:missing_information}

The two measures play different roles in our analysis: the mutual information of Table~\ref{tab:mi_comparison} diagnoses that BFE+MOPED summaries and velocity moments are complementary, while the joint $19$-dimensional Fisher forecast and summary-likelihood emulator (Section~\ref{sec:joint_fisher}) quantify the resulting constraint improvement.

The joint summaries nevertheless do not exhaust the field-level information: adding the velocity moments closes only $11$--$30$ per cent of the BFE+MOPED-to-CFM gap in marginal constraint width, leaving the joint constraints a factor of $1.3$--$2.9$ looser than the particle-level likelihood. This residual gap is not a failure of combining summaries, but identifies phase-space structure that the current linear compression does not capture, and it sets the concrete target for the improved summaries discussed below.

Comparisons with existing MW--LMC mass estimates must account for the fact that different observables probe different aspects of the same system. Stream deflections, timing arguments, LMC internal kinematics, and halo-wake measurements all depend on $M_{\mathrm{LMC}}$, $M_{\mathrm{MW}}$, and the MW potential, but with different parameter combinations and modelling assumptions \citep[e.g.,][]{Erkal2019,Cautun2020,Vasiliev2021,Shipp2021,Fushimi2024}. Our comparison therefore focuses not on matching a single mass estimate, but on how much phase-space information is available, how much is retained by interpretable summaries, and which density and velocity channels carry that information.

Future improvements should preserve this interpretability while adapting the summaries to observational catalogues. Within the BFE framework, this means targeted additions such as selection-function-aware moments, per-tracer weights in place of the equal-mass particle sums used here, anisotropy and tracer-population parameters, or flexible likelihood models on physically grouped coefficient blocks.

\section{Conclusions}
\label{sec:conclusions}

The recent infall of the LMC has driven the MW outer halo into dynamical disequilibrium that encodes the masses and structural parameters of both galaxies in the 6D phase-space distribution of halo tracers. We compare a field-level benchmark, physically interpretable BFE+MOPED summaries, standard all-sky velocity moments, and their joint summary vector for the parameter vector $\boldsymbol{\theta} = (M_{\mathrm{MW}}, M_{\mathrm{LMC}}, c, q)$. Our main conclusions are:

\begin{itemize}
\item For 5,000 tracers between 30 and 120~kpc, inference from the full 6D phase-space distribution gives marginal constraints that are factors of $2.5$--$9.9$ tighter than constraints from 15 all-sky velocity moments (Section~\ref{sec:cfm_full_phase_space_benchmark}).

\item Four BFE+MOPED summaries produce constraints that are factors of $1.2$--$3.3$ tighter than the velocity-moment constraints. Their density and velocity channels show contributions from the reflex dipole, dynamical-friction wake, halo-shape quadrupole, and velocity-dispersion monopoles (Sections~\ref{sec:MOPED}, \ref{sec:fisher_comparison}).

\item A mutual-information analysis shows that BFE+MOPED summaries and standard velocity moments are not redundant. Combining them into the joint $19$-dimensional vector tightens the marginal constraints by up to $15$ per cent relative to BFE+MOPED alone and by $30$--$71$ per cent relative to velocity moments alone, coming within a factor of $1.3$--$2.9$ of the field-level benchmark; the remaining gap quantifies the margin left for future summary designs (Section~\ref{sec:joint_constraints}).

\item Validation on held-out isotropic simulations supports the joint-summary likelihood emulator, with a small marginal coverage deviation for $q$. A separate BFE+MOPED-only stress test with radially biased tracers identifies anisotropy as a leading systematic. The $s_{M_{\mathrm{LMC}}}$ is least affected, while the inferred $M_{\mathrm{MW}}$ and $c$ shift through the second-moment velocity channels (Section~\ref{sec:summary_validation}; Appendix~\ref{app:sbi_validation}).
\end{itemize}

An application to H3, SDSS-V, or a comparable outer-halo catalogue would require survey-specific mock catalogues that include the angular selection function, distance and velocity uncertainties, tracer density profiles, and contamination. The inference model should also include nuisance parameters for the amplitude and radial form of tracer anisotropy $\beta(r)$. After these tests, the joint density and velocity summaries can be used to constrain the MW and LMC masses and the MW halo concentration and flattening. The channel decomposition permits direct assessment of which spatial and kinematic measurements contribute to each inferred parameter.

\section*{Acknowledgements}

We are grateful to Nico Garavito-Camargo for helpful discussions. YST acknowledges support from NSF Grant AST2406729 and a Humboldt Research Award from the Alexander von Humboldt Foundation. X.-X.X. acknowledges the support from the National Key Research and Development Program of China No. 2024YFA1611902, National Natural Science Foundation of China (NSFC) No. 12588202, CAS Project for Young Scientists in Basic Research grant No. YSBR-062, the Strategic Priority Research Program of Chinese Academy of Sciences grant No. XDB1160102 and grant No. CMS-CSST-2025-A11. We further acknowledge the high-performance computing resources provided by the Australian National Computational Infrastructure (grant y89) through the National and ANU Computational Merit Allocation Schemes.

Large-language-model coding assistants (Anthropic Claude Opus 4.7, OpenAI GPT-5.5) were used during this work to refactor analysis scripts, assist with neural-network fine-tuning, and polish the manuscript draft. The authors reviewed and edited all generated content and take full responsibility for the final manuscript.

\section*{Data Availability}

The $N$-body simulation suite developed in this work, HaloDance, will be made publicly available via GitHub at \url{https://github.com/Yanjun-Sheng/HaloDance}. The release will include 101 snapshots spanning the past 2~Gyr, which can be used for stellar-stream modelling, halo-kinematic studies, and forward modelling of Gaia observations.

\appendix

\section{Conditional Flow Matching model details}
\label{app:cfm_details}

The Conditional Flow Matching (CFM) model uses optimal-transport conditional flow matching (OT-CFM) to map a standard Gaussian prior distribution at $t = 0$ to the target parameter-dependent phase-space distribution of the Milky Way (MW) halo dark matter tracer particles at $t = 1$.

For each particle, the input feature vector is $\mathbf{x} = (x, y, z, v_x, v_y, v_z)$, representing Galactocentric positions and velocities, augmented with the Galactocentric radius $r = \sqrt{x^2 + y^2 + z^2}$ and the Galactocentric speed $s = \sqrt{v_x^2 + v_y^2 + v_z^2}$ to incorporate rotation-invariant spatial and velocity information. Coordinates are normalized using standardizing scaling factors derived from the 95th-percentile of the training set:
\begin{equation}
\tilde{\mathbf{x}} = \left( \frac{(x, y, z)}{s_{\mathrm{pos}}},\, \frac{(v_x, v_y, v_z)}{s_{\mathrm{vel}}} \right),
\end{equation}
where $s_{\mathrm{pos}}$ and $s_{\mathrm{vel}}$ are the 95th-percentile radius and speed respectively.

The vector field $\mathbf{u}_t(\mathbf{x} \mid \boldsymbol{\theta})$ is parameterised by a neural network. The time variable $t \in [0, 1]$ is represented via sinusoidal Fourier embeddings with 16 log-spaced frequencies spanning $[1, 1000]$, producing a 32-dimensional embedding vector $\boldsymbol{\gamma}(t) = [\sin(2\pi f_k t),\, \cos(2\pi f_k t)]_{k=1}^{16}$. The physical parameters $\boldsymbol{\theta} = (M_{\mathrm{MW}}, M_{\mathrm{LMC}}, c, q)$ are first standardized, then passed through a two-layer Multilayer Perceptron (MLP) with layer widths 4 $\to$ 128 $\to$ 256 and GELU activation to output a 256-dimensional conditioning vector. The core network is a seven-layer MLP with 1,024 hidden units per layer and GELU activations. It takes the concatenated normalized coordinates, time embeddings, and parameter embeddings as input, and outputs a 6D velocity vector $\mathbf{u}_t(\mathbf{x} \mid \boldsymbol{\theta})$. The total number of parameters is approximately $8.5 \times 10^6$.

The OT-CFM straight-line probability path is
\begin{equation}
p_t(\mathbf{x} \mid \mathbf{x}_1) = \mathcal{N}\bigl(\mathbf{x} \mid t\,\mathbf{x}_1,\; (1 - t)^2\,\mathbf{I}\bigr),
\end{equation}
with constant target vector field
\begin{equation}
\mathbf{u}_t(\mathbf{x} \mid \mathbf{x}_1) = \frac{\mathbf{x}_1 - \mathbf{x}}{1 - t}.
\end{equation}
Training minimises the conditional flow-matching loss over parameter-space coordinates,
\begin{equation}
\mathcal{L}_{\mathrm{CFM}} = \frac{\sum_i w_i\, \left\Vert \mathbf{u}_t(\mathbf{x}_t^{(i)} \mid \boldsymbol{\theta}^{(i)}) - (\mathbf{x}_1^{(i)} - \mathbf{x}_0^{(i)}) \right\Vert^2}{\sum_i w_i},
\end{equation}
where $\mathbf{x}_0 \sim \mathcal{N}(\mathbf{0}, \mathbf{I})$, $\mathbf{x}_1$ is a particle coordinate vector from the training set, $t \sim \mathcal{U}(\varepsilon, 1-\varepsilon)$ (with $\varepsilon = 10^{-3}$), and $\mathbf{x}_t = (1-t)\mathbf{x}_0 + t\mathbf{x}_1$. The sample-specific weights $w_i$ implement a curriculum where the loss initially concentrates on parameter points close to the fiducial cosmology, then smoothly broadens to include the whole parameter space.
Training uses AdamW with learning rate $10^{-4}$, weight decay $10^{-4}$, and batch size 512 for $1.5 \times 10^6$ steps. The learning rate follows a linear warmup over $5 \times 10^4$ steps, remains constant until step $5 \times 10^5$, and cosine decays to zero. Gradients are clipped to 1.0.

To compute the log-likelihood of a single particle $\mathbf{x}$ at parameter point $\boldsymbol{\theta}$, we integrate the probability flow Ordinary Differential Equation (ODE) backward from $t=1$ to $t=0$,
\begin{equation}
\frac{\mathrm{d}}{\mathrm{d}t} \begin{bmatrix} \mathbf{x}_t \\ \log p_t \end{bmatrix} = \begin{bmatrix} \mathbf{u}_t(\mathbf{x}_t \mid \boldsymbol{\theta}) \\ -\operatorname{div}(\mathbf{u}_t)(\mathbf{x}_t \mid \boldsymbol{\theta}) \end{bmatrix}, \qquad \begin{bmatrix} \mathbf{x}_1 \\ \log p_1 \end{bmatrix} = \begin{bmatrix} \tilde{\mathbf{x}} \\ 0 \end{bmatrix}.
\end{equation}
The divergence $\operatorname{div}(\mathbf{u}_t)$ is approximated using the Hutchinson trace estimator with $n_{\mathrm{H}} = 8$ Gaussian probe vectors. The physical coordinate log-likelihood is then
\begin{equation}
\log p(\mathbf{x} \mid \boldsymbol{\theta}) = \log p_0(\mathbf{x}_0) - \int_0^1 \operatorname{div}(\mathbf{u}_t)\,\mathrm{d}t - 3\log s_{\mathrm{pos}} - 3\log s_{\mathrm{vel}}.
\end{equation}
Integration uses a fixed-step RK4 ODE solver with 128 steps.

\section{BFE and MOPED compression details}
\label{app:moped_details}

This appendix gives the construction of the linear MOPED projection matrix $\mathbf{Q}_{\mathrm{score}}$ mapping high-dimensional Basis Function Expansion (BFE) coefficient vectors $\mathbf{f} \in \mathbb{R}^{D}$ to the four compressed BFE+MOPED summaries $\mathbf{m} \in \mathbb{R}^{P}$, with $P=4$.

Before compression, features are standardized using a diagonal covariance approximation. Using $N_{\mathrm{cov}} = 2{,}000$ independent realisations of the MW halo at the fiducial parameter point $\boldsymbol{\theta}_\star = (0.7 \times 10^{12}\,\mathrm{M}_\odot, 1.5 \times 10^{11}\,\mathrm{M}_\odot, 9.415, 1.0)$, we compute the per-feature mean vector $\bar{\mathbf{f}}_\star$ and standard deviation vector $\boldsymbol{\sigma}_\star$ for each of the ten channels. The standardized features are
\begin{equation}
\tilde{\mathbf{f}} = \frac{\mathbf{f} - \bar{\mathbf{f}}_\star}{\boldsymbol{\sigma}_\star}.
\end{equation}
This per-feature scaling sets the diagonal entries of the covariance matrix to unity. The MOPED construction below then treats the remaining off-diagonal terms as negligible, so that $\mathbf{C}\approx\mathbf{I}$ and the MOPED weight vectors reduce to the scores of the standardized means. We adopt this diagonal approximation for numerical stability: with $D\sim10^{4}$ features and $N_{\mathrm{cov}}=2{,}000$ realisations, the full $D\times D$ covariance is rank-limited and cannot be reliably inverted. The choice is conservative rather than lossless. The neglected off-diagonal correlations are individually small (median absolute value $0.015$, 95th percentile $0.046$, 99th percentile $0.066$, with a maximum of $0.99$ from a few strongly coupled modes), but in invertible lower-dimensional subsets including them changes the four-parameter Fisher volume by a factor of $0.15$--$0.51$, so some information is left unrealised.

To test whether this information can be recovered in the full feature space, we regularise the covariance as $\mathbf{C}_\alpha=\alpha\mathbf{I}+(1-\alpha)\mathbf{S}$, where $\mathbf{S}$ is the sample covariance of the standardized features over the $N_{\mathrm{cov}}$ fiducial realisations. We choose $\alpha$ by two-fold cross-validation, recomputing $\mathbf{S}$ and the MOPED weights on one half of the realisations and evaluating the resulting constraints on the held-out half, so that values of $\alpha$ that merely fit sampling noise in $\mathbf{S}$ are penalised. The best cross-validated point, $\alpha=0.9$, tightens the four marginal constraints by factors of $0.97$, $0.97$, $0.99$, and $0.93$ and reduces the Fisher volume by $13$ per cent, while the median ratio to the CFM benchmark remains $2.44$; more aggressive off-diagonal weighting appears tighter in-sample but degrades under cross-validation, indicating covariance overfitting. We therefore retain the diagonal covariance for the production summaries.

The score matrix $\mathbf{D} = [\tilde{\boldsymbol{\mu}}_{,1} \mid \cdots \mid \tilde{\boldsymbol{\mu}}_{,P}] \in \mathbb{R}^{D \times P}$ contains the gradients of the mean standardized feature vector with respect to each parameter $\theta_\alpha$. The derivatives are evaluated by central finite differences,
\begin{equation}
\tilde{\boldsymbol{\mu}}_{,\alpha} \approx \frac{\langle\tilde{\mathbf{f}}\rangle_{\boldsymbol{\theta}_\star + \Delta\theta_\alpha\,\hat{\mathbf{e}}_\alpha} - \langle\tilde{\mathbf{f}}\rangle_{\boldsymbol{\theta}_\star - \Delta\theta_\alpha\,\hat{\mathbf{e}}_\alpha}}{2\Delta\theta_\alpha},
\end{equation}
where $\hat{\mathbf{e}}_\alpha$ is the unit vector in the direction of parameter $\theta_\alpha$ and $\Delta\theta_\alpha$ is the step size. Rather than running new $N$-body simulations at these perturbed coordinates, we evaluate the mean feature vectors by averaging over $N_{\mathrm{deriv}} = 100$ independent realizations drawn from the trained CFM emulator, which interpolates smoothly across the discrete simulation grid.

With this diagonal covariance approximation ($\mathbf{C} \approx \mathbf{I}$), the Gram--Schmidt orthogonalisation of the score directions is computed directly via QR-decomposition of the score matrix,
\begin{equation}
\mathbf{D} = \mathbf{Q}_{\mathrm{score}}\mathbf{R},
\end{equation}
where $\mathbf{Q}_{\mathrm{score}} \in \mathbb{R}^{D \times P}$ is an orthonormal matrix whose columns are orthogonal under the identity metric, and $\mathbf{R} \in \mathbb{R}^{P \times P}$ is upper-triangular. The compressed summaries are then
\begin{equation}
\mathbf{m} = \mathbf{Q}_{\mathrm{score}}^{T}\bigl(\tilde{\mathbf{f}} - \tilde{\boldsymbol{\mu}}_\star\bigr),
\end{equation}
where $\tilde{\boldsymbol{\mu}}_\star \equiv \langle\tilde{\mathbf{f}}\rangle_{\boldsymbol{\theta}_\star}$ is the mean standardized feature vector at the fiducial point. A first-order Taylor expansion at parameters near the fiducial point gives
\begin{equation}
\langle\mathbf{m}\rangle \approx \mathbf{Q}_{\mathrm{score}}^T\mathbf{D}\,\delta\boldsymbol{\theta} = \mathbf{R}\,\delta\boldsymbol{\theta},
\end{equation}
which links the summaries directly to the parameter deviations $\delta\boldsymbol{\theta} = \boldsymbol{\theta} - \boldsymbol{\theta}_\star$.

\section{Variational Mutual Information estimator details}
\label{app:mi_details}

This appendix gives the technical details of the variational Mutual Information (MI) estimator used to compare different summary representations.

The mutual information $I(\boldsymbol{\theta}; \mathbf{s})$ is estimated via the Barber--Agakov variational lower bound \citep{Barber2003,Poole2019},
\begin{equation}
I(\boldsymbol{\theta}; \mathbf{s}) \ge \mathbb{E}_{p(\boldsymbol{\theta},\mathbf{s})}\bigl[\log q_\phi(\boldsymbol{\theta}\mid \mathbf{s})\bigr] + h(\boldsymbol{\theta}),
\end{equation}
where $h(\boldsymbol{\theta}) = -\mathbb{E}_{p(\boldsymbol{\theta})}[\log p(\boldsymbol{\theta})]$ is the analytic prior entropy. For the uniform prior adopted, the entropy is $h(\boldsymbol{\theta}) = \sum_{\alpha} \log(\theta_{\alpha,\mathrm{max}} - \theta_{\alpha,\mathrm{min}})$. The bound is tight in the limit where the variational posterior $q_\phi(\boldsymbol{\theta}\mid \mathbf{s})$ approaches the true posterior $p(\boldsymbol{\theta}\mid \mathbf{s})$.

The variational posterior $q_\phi(\boldsymbol{\theta}\mid \mathbf{s})$ is modelled as a conditional flow-matching posterior over the 4D parameter space. Three independent posteriors are trained: (i) $q_\phi(\boldsymbol{\theta}\mid\mathbf{m})$ conditioned on the 4 BFE+MOPED summaries $\mathbf{m}$, (ii) $q_\phi(\boldsymbol{\theta}\mid\mathbf{v})$ conditioned on the 15 velocity moments $\mathbf{v}$, and (iii) $q_\phi(\boldsymbol{\theta}\mid\mathbf{m},\mathbf{v})$ conditioned on the 19 joint summaries. The training dataset uses the same Latin-hypercube parameter grid split as the MDN-style Gaussian likelihood emulator, consisting of $1{,}663$ training points and $185$ held-out validation points. For each parameter point, $500$ independent $5{,}000$-particle realisations are used to calculate the respective summaries. The vector field $u_\phi(\mathbf{x}_t, t, \mathbf{c}_n)$ parameterising the variational posterior flow is a four-layer MLP with a hidden layer width of 512 and SiLU activations, where $\mathbf{c}_n$ is the conditioning summary vector for the $n$-th training example. The scalar time $t$ is encoded using a sinusoidal Fourier embedding of 16 frequencies. The training loss minimises:
\begin{equation}
\mathcal{L}_{\mathrm{CFM}} = \mathbb{E}\!\left[\,\bigl\Vert u_\phi(\mathbf{x}_t, t, \mathbf{c}_n) - (\boldsymbol{\theta}_n - \mathbf{z})\bigr\Vert_2^2\,\right],
\end{equation}
with $\mathbf{z}\sim\mathcal{N}(0, \mathbf{I})$, $t\sim\mathcal{U}(\varepsilon, 1-\varepsilon)$ (with $\varepsilon = 10^{-3}$), and $\mathbf{x}_t = (1-t)\mathbf{z} + t\boldsymbol{\theta}_n$. Standardized parameter coordinates are used for target flows. We optimize the network using AdamW with a learning rate of $10^{-4}$ and weight decay of $10^{-4}$, with a batch size of 512, for up to $7.5\times 10^5$ steps. The model checkpoint with the minimum validation loss is selected.

To evaluate the lower bound, we compute the log-posterior density of the validation samples under the trained $q_\phi$ using backward probability flow integration. Because the parameter space is low-dimensional ($D = 4$), the vector-field divergence is computed exactly by summing the diagonal entries of the Jacobian matrix, evaluated using reverse-mode automatic differentiation,
\begin{equation}
\operatorname{div}(\mathbf{u}_t) = \sum_{k=1}^4 \frac{\partial u_{t,k}}{\partial x_k}.
\end{equation}
This exact calculation avoids the stochastically noisy Hutchinson trace estimator. The backward integration is solved using a standard Euler solver with 1024 steps.

\section{Summary-likelihood emulator details}
\label{app:sbi_details}

The MDN-style emulator takes the four physical parameters scaled to the unit prior hypercube as input. A shared three-layer trunk with 256 hidden units per layer, SiLU activations, and dropout 0.02 feeds two heads. One predicts the standardised summary mean, and the other predicts the $d(d+1)/2$ entries of a lower-triangular Cholesky factor $\mathbf{L}_{\rm NN}$, where $d$ is the summary dimension ($d=19$ for the joint vector and $d=4$ for BFE+MOPED alone). The covariance is reconstructed as $\boldsymbol{\Sigma}_{\rm NN}=\mathbf{L}_{\rm NN}\mathbf{L}_{\rm NN}^{\top}$, with the diagonal entries passed through a softplus activation and a $10^{-4}$ floor to enforce positive definiteness. The network is trained by minimising
\begin{align}
\mathcal{L} ={}& \mathrm{KL}\!\bigl[\,\mathcal{N}(\hat{\boldsymbol{\mu}}, \hat{\boldsymbol{\Sigma}})\,\Vert\,\mathcal{N}(\boldsymbol{\mu}_{\rm NN}, \boldsymbol{\Sigma}_{\rm NN})\bigr] \notag\\
&+ \lambda\,\bigl\|\log\!\mathrm{diag}(\boldsymbol{\Sigma}_{\rm NN}) - \log\!\mathrm{diag}(\hat{\boldsymbol{\Sigma}})\bigr\|^{2},
\label{eq:nn_loss}
\end{align}
with $\lambda=0.1$. Training uses AdamW with learning rate $10^{-3}$, weight decay $10^{-4}$, and batch size 256 on 1,663 training parameter points, with early stopping on the held-out validation loss.

To propagate emulator-training uncertainty, we train an ensemble of five networks with independent random seeds and combine their predictions at inference time. The effective likelihood uses the ensemble-averaged mean and a covariance that includes both the predicted summary covariance and the scatter among network mean predictions,
\begin{align}
\boldsymbol{\mu}_{\rm eff}(\boldsymbol{\theta})
&= \frac{1}{M}\!\sum_{m=1}^{M} \boldsymbol{\mu}_m(\boldsymbol{\theta}), \notag\\
\boldsymbol{\Sigma}_{\rm eff}(\boldsymbol{\theta})
&= \frac{1}{M}\!\sum_{m=1}^{M} \boldsymbol{\Sigma}_m(\boldsymbol{\theta})
+ \mathrm{diag}\!\bigl[\mathrm{Var}_m\,\boldsymbol{\mu}_m(\boldsymbol{\theta})\bigr].
\label{eq:nn_ensemble_cov}
\end{align}
The first covariance term represents the conditional summary covariance at fixed $\boldsymbol{\theta}$, and the second term adds a diagonal emulator-uncertainty floor. We retain only the diagonal of the between-network covariance, because off-diagonal entries estimated from just $M=5$ members would inject noisy correlations; the restriction conservatively inflates the marginal variances. This floor is small, contributing a median of $0.3$ per cent of the summary variance across the validation set and at most about $2$ per cent, so we use the deterministic five-member ensemble in production without additional per-call predictive resampling.

\section{Validation and calibration of the summary SBI pipeline}
\label{app:sbi_validation}

This appendix describes the calibration and accuracy tests for the summary-level Simulation-Based Inference (SBI) pipeline.

\begin{figure*}
    \centering
    \begin{minipage}{0.48\textwidth}
        \centering
        \includegraphics[width=\textwidth]{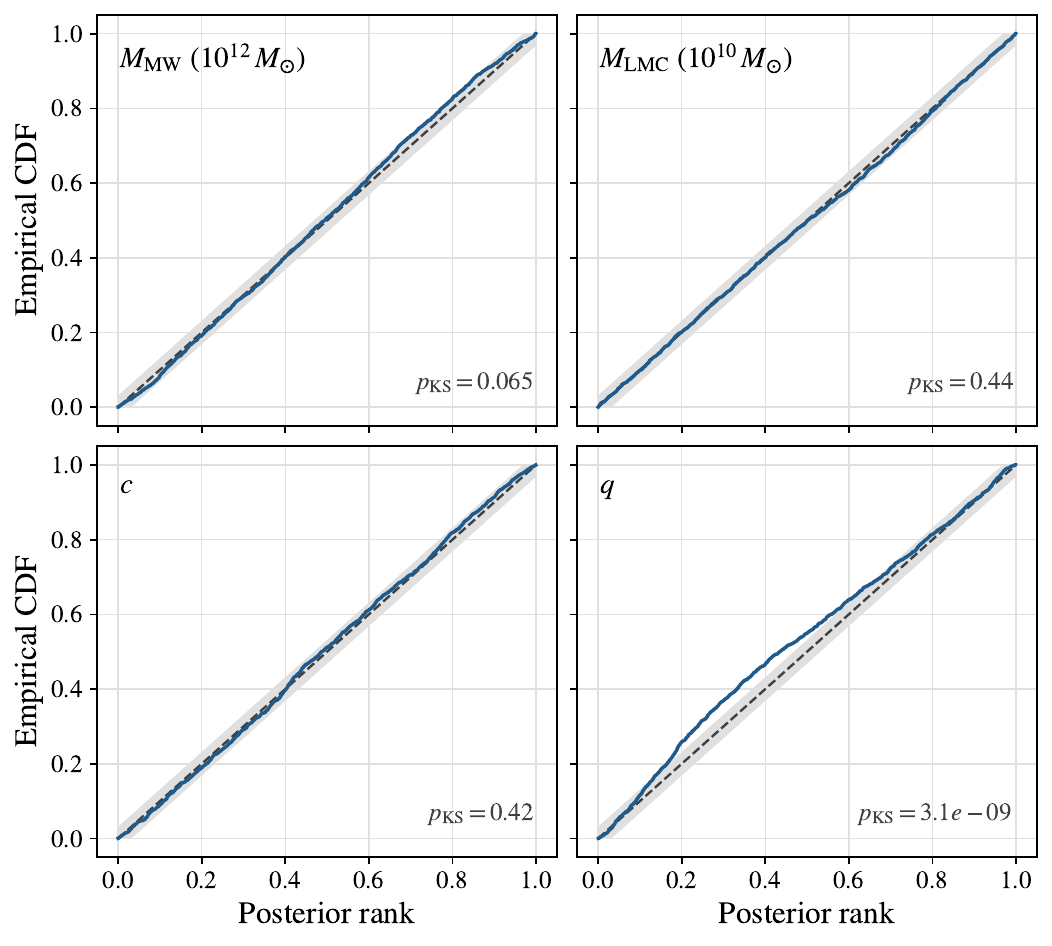}
    \end{minipage}
    \hfill
    \begin{minipage}{0.48\textwidth}
        \centering
        \includegraphics[width=\textwidth]{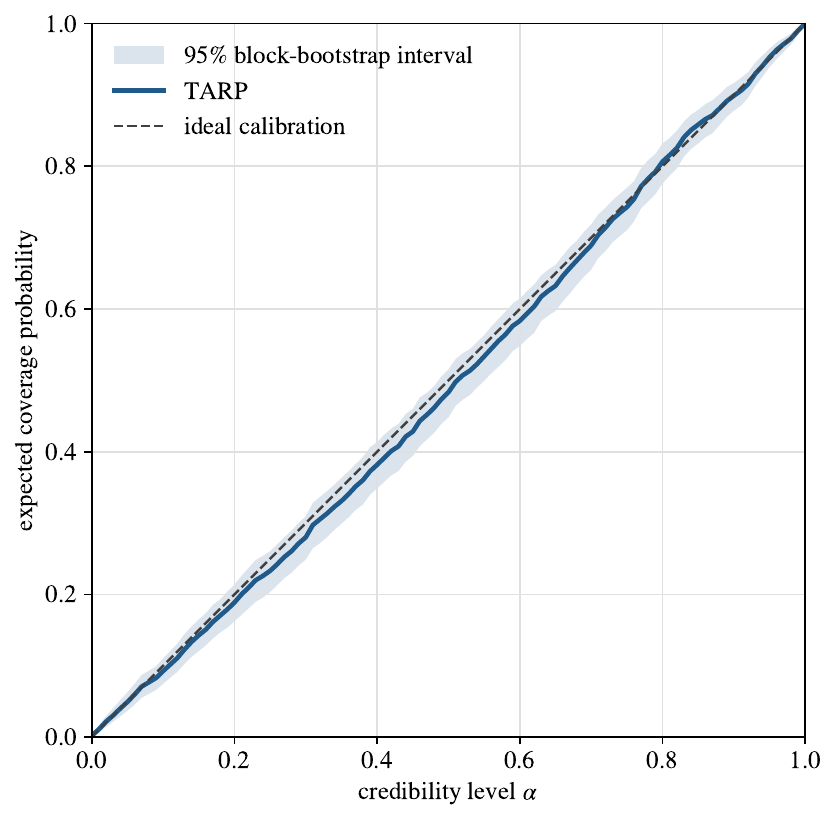}
    \end{minipage}
    \vspace{0.5em}

    \includegraphics[width=0.86\textwidth]{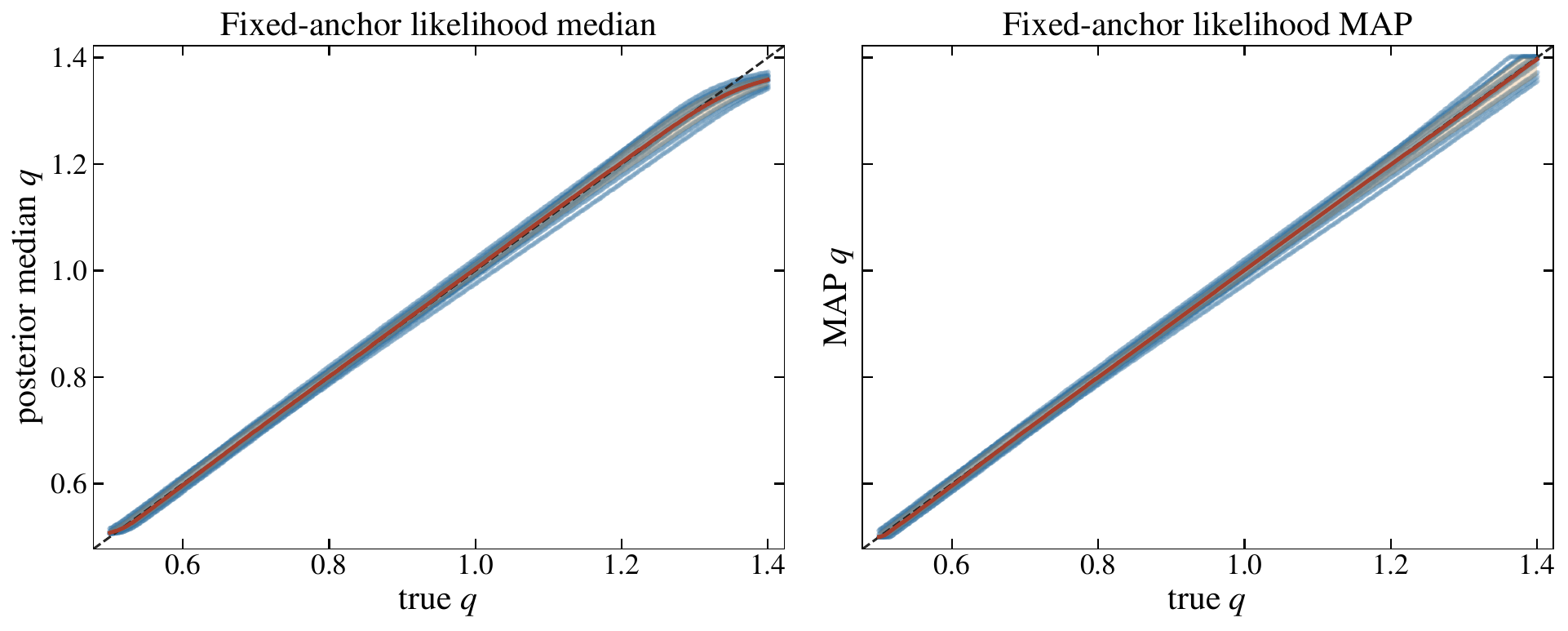}
    \caption{Validation diagnostics for the MDN-style Gaussian summary-likelihood emulator. The top-left panel shows marginal posterior P-P ranks on the validation set, with dashed diagonals for calibrated marginals and grey bands for 95\% acceptance regions. The top-right panel shows joint TARP coverage, with the light blue band marking block-bootstrap uncertainty. The bottom panels show the dense $q$ scan at the fiducial anchor, where both the posterior median and likelihood maximum track the one-to-one relation smoothly. The faint lines are the individual repeated realisations at each scanned $q$, and the solid line is their mean.}
    \label{fig:validation_diagnostics}
\end{figure*}

We validate the posterior calibration of the individual parameters by evaluating their marginal percentile--percentile (P-P) ranks. For each parameter $\theta_\alpha$ and a validation dataset of repeated realisations at known parameter values, we compute the posterior rank of the true value,
\begin{equation}
r_\alpha = \frac{1}{N_{\mathrm{samp}}} \sum_{j=1}^{N_{\mathrm{samp}}} \mathbb{I}\!\left[ \theta_\alpha^{(j)} < \theta_{\alpha,\mathrm{true}} \right],
\label{eq:pp_rank_definition}
\end{equation}
where $\theta_\alpha^{(j)}$ is the value of the parameter in the $j$-th posterior sample, $\theta_{\alpha,\mathrm{true}}$ is the true parameter value, and $\mathbb{I}$ is the indicator function. A calibrated marginal posterior yields ranks that are uniformly distributed between 0 and 1. The P-P curve is the empirical cumulative distribution function (CDF) of these ranks. The top-left panel of Figure~\ref{fig:validation_diagnostics} shows the marginal P-P plots for $(M_{\mathrm{MW}}, M_{\mathrm{LMC}}, c, q)$. The Kolmogorov--Smirnov (KS) test values are $p_{\mathrm{KS}} = 0.065$, $0.44$, and $0.42$ for $M_{\mathrm{MW}}$, $M_{\mathrm{LMC}}$, and $c$, indicating statistical consistency with perfect calibration. The $q$ parameter shows a slight departure from uniformity ($p_{\mathrm{KS}} = 3.1 \times 10^{-9}$), reflecting marginal coverage errors, though the absolute deviation remains small.

To assess whether the posteriors capture the joint parameter correlations, we apply the TARP diagnostic \citep{Lemos2023}. TARP measures the joint coverage by calculating the posterior mass inside a Euclidean sphere centred on an independent random point $\boldsymbol{\theta}_{\mathrm{ref}}$, with a radius equal to the distance from $\boldsymbol{\theta}_{\mathrm{ref}}$ to the true parameter $\boldsymbol{\theta}_{\mathrm{true}}$,
\begin{equation}
\alpha_{\mathrm{TARP}} = \frac{1}{N_{\mathrm{samp}}} \sum_{j=1}^{N_{\mathrm{samp}}} \mathbb{I}\!\left[ \left\|\boldsymbol{\theta}^{(j)} - \boldsymbol{\theta}_{\mathrm{ref}}\right\| < \left\|\boldsymbol{\theta}_{\mathrm{true}} - \boldsymbol{\theta}_{\mathrm{ref}}\right\| \right].
\label{eq:tarp_definition}
\end{equation}
For a calibrated joint posterior, the distribution of $\alpha_{\mathrm{TARP}}$ values across the validation set is uniform. The top-right panel of Figure~\ref{fig:validation_diagnostics} shows that the empirical TARP curve matches the diagonal within the 95\% bootstrap uncertainty, indicating that the joint four-dimensional posterior passes this coverage test.

We evaluate the point-estimate accuracy using True-vs-Predicted validation grids. Figure~\ref{fig:true_vs_predicted} shows the posterior medians and 16th-to-84th percentile uncertainties against the true parameters. The median absolute residuals are $0.15 \times 10^{11}\,\mathrm{M}_\odot$ for $M_{\mathrm{MW}}$, $0.79 \times 10^{10}\,\mathrm{M}_\odot$ for $M_{\mathrm{LMC}}$, $0.38$ for $c$, and $0.015$ for $q$. These residuals are smaller than the median 68\% posterior widths ($0.23$, $1.21$, $0.63$, and $0.020$, respectively), indicating that the point estimates recover the true values without systematic offsets.

The marginal posterior medians of the flattening parameter $q$ show a small over-clustering at recurring locations in the P-P ranks and point estimates. To determine whether this crowding is caused by discontinuities in the emulator likelihood, we evaluate the MDN-style Gaussian likelihood ensemble for the joint 19-dimensional summaries on a dense grid scan of $q$ at the fiducial point, varying $q_{\mathrm{true}}$ from $0.5$ to $1.4$ in steps of $10^{-3}$. The bottom panels of Figure~\ref{fig:validation_diagnostics} show that both the likelihood maximum $q_{\mathrm{MAP}}$ and the posterior median follow the diagonal smoothly. This rules out discontinuities in the emulator likelihood. The same clustering appears when the validation is repeated with the BFE+MOPED summaries or the velocity moments alone, so it is not introduced by the joint-summary construction. Instead, the clustering reflects parts of the prior where the summaries have weak leverage on $q$, so the marginal posterior is shaped mainly by the prior and by degeneracies with the other parameters.

\FloatBarrier

\bibliographystyle{mnras}
\bibliography{Manuscript}

\bsp
\label{lastpage}
\end{document}